\documentclass[12pt]{article}
\usepackage{epsfig,makeidx,color}
\usepackage{amsbsy}
\usepackage{amsmath}
\usepackage{amssymb}
\usepackage{citesort}
\newcounter{mytempeqncnt}

\begin{document}

\title{\vspace{-2.0cm} \bf Golden Space-Time Trellis Coded Modulation}

\author{Yi Hong, Emanuele Viterbo, and Jean-Claude Belfiore
\hspace{0.2cm}\thanks{\scriptsize  Yi Hong is with the Institute
for Telecommunications Research, University of South Australia,
Australia, Emanuele Viterbo is with Politecnico di Torino, Italy,
Jean-Claude Belfiore is with ENST, Paris, France. E-mail :
yi.hong@unisa.edu.au,viterbo@polito.it,belfiore@enst.fr. } }

\maketitle
\date
\vspace{-1cm}
\begin{abstract}
In this paper, we present a multidimensional trellis coded
modulation scheme for a high rate $2\times 2$ multiple-input
multiple-output system over slow fading channels. Set partitioning
of the Golden code \cite{Golden05} is designed specifically to
increase the minimum determinant. The branches of the outer trellis
code are labeled with these partitions and Viterbi algorithm is
applied for trellis decoding. In order to compute the branch metrics
a sphere decoder is used. The general framework for code design and
optimization is given. Performance of the proposed scheme is
evaluated by simulation and it is shown that it achieves significant
performance gains over uncoded Golden code.
\end{abstract}
\vspace{-1cm}
\paragraph{Index terms:} Lattice, set partitioning, trellis coded
modulation, Golden code, diversity, coding gain, minimum
determinant.
\baselineskip 18pt


\section{Introduction}
Space-time codes were proposed in \cite{Tarokh98} as a combination
of channel coding with transmit diversity techniques in order to
enhance data rates and reliability in multi-antenna wireless
communications systems. In the coherent scenario, where the channel
state information (CSI) is available at the receiver, the design
criteria for space-time codes in slow fading channels were
developed: {\em rank} and {\em determinant criteria}
\cite{Tarokh98}. The design criteria aim to maximizing the minimum
rank and determinant of the codeword distance matrix in order to
maximize the diversity and coding gains. This in turn guarantees the
best possible asymptotic slope of the error performance curve on a
log-log scale, as well as a shift to the left of the curve.

Subsequent works resulted new space-time trellis codes, orthogonal
space-time block codes  \cite{Alamouti,STBCTarokh}, etc.  In
particular, orthogonal space-time block codes attracted a lot of
interest due to their low decoding complexity and high diversity
gain. Further work produced full diversity, full rate algebraic
space-time block codes for any number of transmit antennas, using
number theoretical methods \cite{DAST1,DAST2,UniversalSTC03}. A
general family of full rank and full rate linear dispersion
space-time block codes based on cyclic division algebras was
proposed in \cite{Sethuraman03}. However, all the above coding
schemes do not always exploit the full potential of the
multiple-input multiple-output (MIMO) system in terms of
diversity-multiplexing gain trade-off \cite{Tse03}. In
\cite{Golden05}, the Golden code was proposed as a full rate and
full diversity code for $2\times 2$ MIMO systems with non-vanishing
minimum determinant (NVD). It was shown in \cite{Elia05} how this
property guarantees to achieve  the diversity-multiplexing gain
trade-off.

In this work we focus on the slow fading model, where it is assumed
that the channel coefficients are fixed over the duration of a
fairly long frame. In such a case, in order to reduce the decoding
complexity, concatenated coding schemes are appropriate. Space-time
trellis codes (STTCs) transmitting PSK or QAM symbols from each
antenna were designed according to both {\em rank} and {\em
determinant} criteria \cite{Tarokh98}. A more flexible design, using
a concatenated scheme, enables to separate the optimization of the
two design criteria. As an inner code, we can use a simple
space-time block code, which can guarantee full diversity for any
spectral efficiency (e.g. Alamouti code \cite{Alamouti}). An outer
code is then used to improve the coding gain. Essentially two
approaches are available:
\begin{enumerate}
\item   bit-interleaved coded modulation (BICM) using a powerful
binary code and computing bit reliability (soft outputs) for the
inner code;
\item   trellis coded modulation (TCM) using set partitioning of the inner code.
\end{enumerate}
The first approach requires a soft output decoder of the inner code,
which can have high complexity as the spectral efficiency increases.
The second approach, considered in this paper, overcomes above
limitations and is appropriate for high data rate systems. We note
how the NVD property for the inner code is essential when using a
TCM scheme: such schemes usually require a constellation expansion,
which will not suffer from a reduction of the minimum determinant.
This advantage is not available with Super-orthogonal space-time
trellis codes proposed in \cite{Jafarkhani03}.

A first attempt to concatenate the Golden code with an outer
trellis code was made in \cite{David05}. Set partitioning of the
inner code was used to increase the minimum determinant of the
inner codewords, which label the branches of the outer trellis
code. The resulting {\em ad hoc} scheme suffered from a high
trellis complexity.

In this paper, we develop  general framework for code design and
optimization for Golden Space-Time Trellis Coded Modulation
(GST-TCM) schemes. In \cite{Calderbank,Ungerboeck,Forney1,Forney2},
lattice set partitioning, combined with a trellis code, is used to
increase the minimum square Euclidean distance between codewords.
Here, it is used to increase the minimum  determinant. The Viterbi
algorithm is used for trellis decoding, where the branch metrics are
computed by using a lattice sphere decoder \cite{Viterbo99} for the
inner code.

We consider partitions of the Golden code with increasing minimum
determinant. In turn, this corresponds to a $\mathbb{Z}^8$ lattice
partition, which is labeled by using a sequence of nested binary
codes. The resulting partitions are selected according to a design
criterion that is  similar to Ungerboeck design rules
\cite{Ungerboeck,Biglieri05}. We design different GST-TCMs and
optimize their performance according to the design criterion.

For example, we will show that 4 and 16 state TCMs achieve
significant performance gains of 3dB and 4.2dB, at frame error rate
(FER) of $10^{-3}$, over the uncoded Golden code at spectral
efficiencies of 7 and 6 bits per channel use (bpcu), respectively.

The rest of the paper is organized as follows. Section 2 introduces
the system model. Section 3 presents a set partitioning of the
Golden code which increases the minimum determinant. Section 4 the
GST-TCM presents design criteria and various examples of our scheme.
Conclusions are drawn in Section 5.

The following notations are used in the paper. Let $T$ denote
transpose and $\dagger$ denote Hermitian transpose. Let
$\mathbb{Z}$, $\mathbb{Q}$, $\mathbb{C}$ and $\mathbb{Z}[i]$ denote
the ring of rational integers, the field of rational numbers, the
field of complex numbers, and the ring of Gaussian integers, where
$i^2 =-1$. Let $GF(2)=\{0,1\}$ denote the binary Galois field. Let
$\mathbb{Q}(\theta )$ denote an algebraic number field generated by
the primitive element $\theta $. The real and imaginary parts of a
complex number are denoted by $\Re{\left( \cdot \right)}$ and
$\Im{\left( \cdot \right)}$. The $m\times m$ dimensional identity
matrix is denoted by $\mathbf{I}_m$. The $m\times n$ dimensional
zero matrix is denoted by $\mathbf{0}_{m\times n}$. The Frobenius
norm of a matrix is denoted by $\|\cdot\|_F$. Let $\mathbb{Z}^{8}$
be the $8$-dimensional integer lattice and let $D_4$ and $E_8$
(Gosset lattice) denote the densest sphere packing in $4$ and $8$
dimensions \cite{Spherepacking}.

\section{System Model}\label{sec:sysmod}

We consider a $2\times 2$ $(n_T = 2, n_R = 2)$ MIMO system over slow
fading channels. The received signal matrix ${\mathbf{Y}} \in
\mathbb{C}^{2\times{2L}}$, where $2L$ is the \ {\em frame} length,
is given by
\begin{equation}
{\mathbf{Y}}={\mathbf{H}}{\mathbf{X}}+{\mathbf{Z}}, \label{SM}
\end{equation}
where ${\mathbf{Z}}\in \mathbb{C}^{2\times {2L}}$ is the complex
white Gaussian noise with i.i.d. samples $\sim
\mathcal{N}_{\mathbb{C}} (0,N_0)$, $\mathbf{H}\in
\mathbb{C}^{2\times {2}}$ is the channel matrix, which is constant
during a frame and varies independently from one frame to another.
The elements of $\mathbf{H}$ are assumed to be i.i.d. circularly
symmetric Gaussian random variables $\sim \mathcal{N}_{\mathbb{C}}
(0,1)$. The channel is assumed to be known at the receiver.

In (\ref{SM}), $\mathbf{X}=[{X}_{1}$,...,${X}_{t},..., {X}_{L}]\in
\mathbb{C}^{2\times {2L}}$ is the transmitted signal matrix, where
${X}_{t} \in \mathbb{C}^{2\times {2}}$. There are three different
options for selecting inner codewords ${X}_{t}, t=1,\ldots,L$:
\begin{enumerate}
    \item ${X}_{t}$ is a codeword of the Golden code $\cal G$, i.e.,
    \begin{equation}
    {X}_{t}=\frac{1}{\sqrt{5}}\left[
    \begin{array}{cc}
    \alpha \left( a+b\theta \right)  & \alpha \left( c+d\theta \right)  \\
    i \bar{\alpha} \left( c+d\bar{\theta} \right)  & \bar{\alpha}
    \left(  a+b\bar{\theta} \right)
    \end{array}%
    \right] , \label{goldencodeword}
    \end{equation}
    where $a,b,c,d \in \mathbb{Z}[i]$ are
    the information symbols, $\theta = 1- \bar{\theta } =
    \frac{1+\sqrt{5}}{2}$, $\alpha = 1 + i -i\theta$,
    $\bar{\alpha} = 1+i(1-\bar\theta )$, and the factor $\frac{1}{\sqrt{5}}$ is necessary
    for energy normalizing purposes \cite{Golden05}.
    \item ${X}_{t}$ are independently selected from a linear subcode of the
    Golden code;
    \item A trellis code is used as the outer code encoding across
    the symbols $X_{t}$, selected from partitions of  $\cal G$.
\end{enumerate}
We denote Case 1 as the {\em uncoded Golden code}, Case 2 as the
{\em Golden subcode}, and Case 3 as the {\em Golden space-time
trellis coded modulation}.

In this paper, we use $Q$--QAM constellations as information symbols
in (\ref{goldencodeword}), where $Q = 2^{\eta}$. We assume the
constellation is scaled to match $\mathbb{Z}[i]+(1+i)/2$, i.e., the
minimum Euclidean distance is set to 1 and it is centered at the
origin. For example, the average energy is $E_s =0.5,1.5,2.5,5,10.5$
for $Q=4$,8,16,32,64. Without loss of generality, we will neglect
the translation vector $(1+i)/2$ and assume the $Q$--QAM
constellation is carved from $\mathbb{Z}[i]$, using a square (or
cross-shaped) bounding region ${\cal B}_{\rm QAM}$, typical for
QAMs. For convenience in our analysis, we will choose ${\cal B}_{\rm
QAM}$ to be in the positive quadrant. In order to minimize the
transmitted energy of this constellation, we center it with by
adding a suitable translation.

Signal to noise ratio is defined as $\text{SNR} = n_T E_{b}/N_{0}$,
where $E_{b} = E_{s}/q$ is the energy per bit and $q$ denotes the
number of information bits per symbol. We have $N_{0} =
2{\sigma}^2$, where ${\sigma}^2$ is the noise variance per real
dimension, which can be adjusted as $\sigma ^{2} =
(n_{T}E_{b}/2)10^{(\text{-SNR}/10)}$.

Assuming that a codeword $\mathbf{X}$ is transmitted, the maximum-likelihood
receiver might decide erroneously in favor of another codeword
$\mathbf{\hat{X}} $. %
Let $r$ denote the rank of the {\em codeword difference matrix}
${\mathbf{X}} -\mathbf{ \hat{X}} $. Since the Golden code is a full
rank code, we have  $r=n_{T}=2$.

Let $\lambda_{j}, j=1,\ldots, r$, be the eigenvalues of the {\em
codeword distance matrix} $\mathbf{A}=( \mathbf{X} -
\mathbf{\hat{X}})({\mathbf{X} -\mathbf{\hat{X}}})^{\dagger}$. Let
$\Delta =\underset{j=1}{\overset{n_{T}}{\prod }}{\lambda }_{j}$ be
the determinant of the codeword distance matrix $\mathbf A$ and
$\Delta_{\min}$ be the corresponding {\em minimum determinant},
which is defined as
\begin{equation}
\Delta _{\min}=\underset{\mathbf{X\neq \hat{X}}}{\min}\det\left(
\mathbf{A}\right).
\end{equation}
The pairwise error probability (PWEP) is upper bounded by
\begin{equation}
P\left( {\mathbf{X} }\rightarrow {\mathbf{\hat{X}} }\right) \leq
\left( \Delta_{\min} \right) ^{-n_{R}}\left(
\frac{E_{s}}{N_{0}}\right) ^{-n_{T} n_{R}} \label{PWEP}
\end{equation}
where $n_{T} n_{R}$ is the {\em diversity gain} and $\left(
\Delta_{min} \right)^{1/{n_{T}}}$ is the {\em coding gain}
\cite{Tarokh98}. In the case of linear codes analyzed in this
paper, we can simply consider the all-zero codeword matrix and we
have
\begin{equation}
\Delta _{\min}=\underset{\mathbf{X} \neq {\bf 0}_{2\times
2L}}{\min}\left|\det\left( \mathbf{X} \mathbf{X}^\dagger
\right)\right|^2.
\end{equation}

In order to compare two coding schemes for the $n_T \times n_R$ MIMO
system, supporting the same information bit rate, but different
minimum determinants (${\Delta_{\min,1}}$ and ${\Delta_{\min,2}}$)
and different constellation energies ($E_{s,1}$ and $E_{s,2}$), we
define the asymptotic coding gain as
\begin{equation} \label{eq:asymcodgain}
\gamma_{as} =
\frac{\sqrt[n_R]{\Delta_{\min,1}}/E_{s,1}}{\sqrt[n_R]{\Delta_{\min,2}}/E_{s,2}}
\end{equation}
We will only consider the case with $n_R=2$, which enables to
exploit the full power of the Golden code with the minimum number of
receive  antenna. Adding extra receive antennas can increase the
receiver diversity and hence performance at the cost of higher
complexity.

Performance of both uncoded Golden code (Case 1) and  Golden subcode
(Case 2) systems can be analyzed for $L=1$. The Golden code
$\mathcal{G}$ has full rate, full rank, and the minimum determinant
is $\delta_{\min} = \frac{1}{5}$ \cite{Golden05}; thus, for Case 1,
$\Delta_{\min}=\delta_{\min}$. For Case 2, a linear subcode of
$\mathcal{G}$ is selected such that $\Delta_{\min}>1/5$. For GST-TCM
(Case 3) we consider $L>1$ and the minimum determinant can be
written as
\begin{equation}
\Delta _{\min }=\min_{\mathbf{X}\neq {\mathbf{0}_{2\times 2L}}} \det
(\mathbf{XX}^{\dagger })=\min_{\mathbf{X}\neq {\mathbf{0}_{2\times 2L}}} \det
\left( \sum_{t=1}^{L}\left( X_{t} X_{t}^{\dagger }\right) \right).
\label{Determinant}
\end{equation}
A code design criterion attempting to maximize $\Delta_{\min}$ is
hard to exploit, due to the non-additive nature of the determinant
metric in (\ref{Determinant}). Since $X_t X_t^{\dagger}$ are
positive definite matrices, we use the following determinant
inequality \cite{matrixbook}:
\begin{equation}
\Delta _{\min } \geq  \min_{\mathbf{X}\neq {\mathbf{0}_{2\times
2L}}} \sum\limits_{t=1}^{L}\det \left( X_{t}X_{t}^{\dagger
}\right) = \Delta_{\min}'. \label{lowerbound}
\end{equation}
The lower bound $\Delta_{\min}^{'}$ will be adopted as the guideline
of our concatenated scheme design. In particular we will design
trellis codes that attempt to maximize $\Delta_{\min}^{'}$, by using
set partitioning to increase the number and the magnitude of non
zero terms $\det \left( X_{t}X_{t}^{\dagger }\right) $ in
(\ref{lowerbound}).

Note that our design criterion is based on the optimization of an
upper bound to the upper bound on the worst case pairwise error
probability in (\ref{PWEP}). Nevertheless, simulation results show
that the codes with the largest $\Delta_{\min}^{'}$ always performed
better.

\section{Uncoded Golden code and its subcodes}\label{sec:BCM}

In both Case 1 and Case 2, the symbols $X_t$ are transmitted
independently in each time slot $t=1,\ldots, L$. The subscript $t$
will be omitted for brevity. We recall below the fundamental
properties of the Golden code deriving from its algebraic structure
\cite{Golden05}.
\begin{itemize}
\item Full-rank: the cyclic division algebra structure guarantees that all the
codewords have full rank (i.e., non zero determinant).
\item Full-rate: the spectral efficiency is of two $Q$--QAM information symbols per channel use,
(i.e., $2\log_2(Q)$ bits/s/Hz) and saturates the two degrees of
freedom of the $2\times2$ MIMO system.
\item Cubic shaping: this relates to the cubic shape of the vectorized
eight-dimensional constellation and guarantees that no shaping loss
is induced by the code.
\item Non-vanishing determinant for increasing  $Q$--QAM size: this
property is derived from the discrete nature of the infinite Golden
code.
\item Minimum determinant $\delta_{\min} = 1/5$: this preserves the coding
gain for any $Q$--QAM size.
\item Achieves the Diversity Multiplexing gain frontier for 2TX-2RX
antennas \cite{Elia05}
\end{itemize}
These particular properties of the Golden code are the key to its
performance improvement over all previously proposed codes. The NVD
property is especially useful for adaptive modulation schemes or
whenever we need to expand the constellation to compensate for a
rate loss caused by an outer code, as in TCM.
\subsection{Uncoded Golden code}
At any time $t$, the received signal matrix $Y =(y_{ij})\in
\mathbb{C}^{2\times 2}$ can be written as
\begin{equation}
Y=\mathbf{H}X+Z, \label{newSM}
\end{equation}
where  ${\mathbf H}=(h_{ij})$ is the channel matrix, $X=(x_{ij})$
the transmitted signal matrix and $Z=(z_{ij})$ the noise matrix.
Vectorizing and separating real and imaginary parts in (\ref{newSM})
yields
\begin{equation}
\mathbf{y} ={\cal H} \mathbf{x} + \mathbf{z}, \label{vector}
\end{equation}
where ${\cal H}$ is given in (\ref{vectorh}) and
\begin{eqnarray}
\mathbf{y}&=&\left[ \Re{\left( y_{11}\right)},\, \Im{\left( y_{11}\right)},\,
\Re{\left( y_{21}\right)},\, \Im{\left( y_{21}\right)},\, \Re{\left(
y_{12}\right)},\, \Im{\left( y_{12}\right)},\, \Re{\left( y_{22}\right)},\,
\Im{\left( y_{22}\right)} \right]^{T} \label{vectory} \\
\mathbf{z}&=&\left[ \Re{\left( z_{11}\right)},\, \Im{\left( z_{11}\right)},\,
\Re{\left( z_{21}\right)},\, \Im{\left( z_{21}\right)},\, \Re{\left(
z_{12}\right)},\, \Im{\left( z_{12}\right)},\, \Re{\left( z_{22}\right)},\,
\Im{\left( z_{22}\right)} \right]^{T} \label{vectorn}\\
\mathbf{x}&=&\left[ \Re{\left( x_{11}\right)},\, \Im{\left( x_{11}\right)},\,
\Re{\left( x_{21}\right)},\, \Im{\left( x_{21}\right)},\, \Re{\left(
x_{12}\right)},\, \Im{\left( x_{12}\right)},\, \Re{\left( x_{22}\right)},\,
\Im{\left( x_{22}\right)} \right]^{T} \label{vectorx}
\end{eqnarray}
\begin{figure*}[!t]
\normalsize \setcounter{mytempeqncnt}{\value{equation}}
\begin{equation}
{\cal H}= \left[
\begin{array}{cccccccc}
\Re{\left( h_{11}\right)}  & -\Im{\left( h_{11}\right)}  &
\Re{\left(h_{12}\right)}  & -\Im{\left( h_{12}\right)}  & 0 & 0 & 0 & 0 \\
\Im{\left( h_{11}\right)}  & \Re{\left( h_{11}\right)}  &
\Im{\left(h_{12}\right)}  & \Re{\left( h_{12}\right)}  & 0 & 0 & 0 & 0 \\
\Re{\left( h_{21}\right)}  & -\Im{\left( h_{21}\right)}  &
\Re{\left(h_{22}\right)}  & -\Im{\left( h_{22}\right)}  & 0 & 0 & 0 & 0 \\
\Im{\left( h_{21}\right)}  & \Re{\left( h_{21}\right)}  &
\Im{\left(h_{22}\right)}  & \Re{\left( h_{22}\right)}  & 0 & 0 & 0 & 0 \\
0 & 0 & 0 & 0 & \Re{\left( h_{11}\right)}  &
-\Im{\left(h_{11}\right)} & \Re{\left( h_{12}\right)}  & -\Im{\left( h_{12}\right)}\\
0 & 0 & 0 & 0 & \Im{\left( h_{11}\right)}  & \Re{\left(
h_{11}\right)} &
\Im{\left( h_{12}\right)}  & \Re{\left( h_{12}\right)}  \\
0 & 0 & 0 & 0 & \Re{\left( h_{21}\right)}  & -\Im{\left(
h_{21}\right)} &
\Re{\left( h_{22}\right)}  & -\Im{\left( h_{22}\right)}  \\
0 & 0 & 0 & 0 & \Im{\left( h_{21}\right)}  & \Re{\left(
h_{21}\right)} & \Im{\left( h_{22}\right)}  & \Re{\left(
h_{22}\right)}
\end{array}%
\right], \label{vectorh}
\end{equation}
\setcounter{equation}{\value{mytempeqncnt}} \hrulefill
\vspace*{4pt}
\end{figure*}
\addtocounter{equation}{1}
Lattice decoding is employed to find $\mathbf{x}$ such that
\begin{equation}
\min_{\mathbf{x} \in \mathbf{R} {\mathbb{Z}^8} }\left\Vert
\mathbf{y}-{\cal H}\mathbf{x}\right\Vert ^{2},
\label{latticedecoding}
\end{equation}
where
\begin{equation}
\mathbf{R} = \frac{1}{\sqrt{5}}\left[
\begin{array}{cccccccc}
1 & -\bar{\theta}  & \theta  & 1 & 0 & 0 & 0 & 0 \\
\bar{\theta}  & 1 & -1 & \theta  & 0 & 0 & 0 & 0 \\
0 & 0 & 0 & 0 & -\theta  & -1 & 1 & -\bar{\theta} \\
0 & 0 & 0 & 0 & 1 & -\theta  & \bar{\theta}  & 1 \\
0 & 0 & 0 & 0 & 1 & -\bar{\theta}  & \theta  & 1 \\
0 & 0 & 0 & 0 & \bar{\theta}  & 1 & -1 & \theta  \\
1 & -\theta  & \bar{\theta}  & 1 & 0 & 0 & 0 & 0 \\
\theta  & 1 & -1 & \bar{\theta}  & 0 & 0 & 0 & 0%
\end{array}%
\right]. \label{rotationmatrix}
\end{equation}
is a rotation matrix preserving the shape of the QAM information symbols
$a,b,c,d$. For this reason we will identify the Golden code with the rotated
lattice $\mathbf{R}\mathbb{Z}^8= \{\mathbf{x}=\mathbf{R}\mathbf{u}\}$ where
\begin{equation}
\mathbf{u}= \left[ \Re{\left( a\right)},\, \Im{\left( a\right)},\,
\Re{\left( b\right)},\, \Im{\left(b\right)},\,
\Re{\left(c\right)},\, \Im{\left( c\right)},\,
\Re{\left(d\right)},\, \Im{\left(d\right)} \right].^{T}
\end{equation}

The lattice decoding problem can be rewritten as
\begin{equation}
\min_{\mathbf{u} \in  {\mathbb{Z}^8} }\left\Vert \mathbf{y}-{\cal
H}\mathbf{R}\mathbf{u}\right\Vert ^{2}~.
\end{equation}

\begin{table}[t]
\begin{center}
\begin{tabular}{|c|c|c|c|c|} \hline
\rule{0mm}{6mm} $k$ & Golden subcode & Lattice & Binary code & $\Delta_{\min}$ \\
\hline\hline
\rule{0mm}{6mm} 0 & $\mathcal{G}$   & $\mathbb{Z}^8$   & $C_0 = (8,8,1)$ & $\delta_{\min}$\\
\rule{0mm}{6mm} 1 & $\mathcal{G}_1$ & $D_4^2$ & $C_1 = (8,6,2)$ & $2\delta_{\min}$\\
\rule{0mm}{6mm} 2 & $\mathcal{G}_2$ & $E_8$   & $C_2 = (8,4,4)$ & $4\delta_{\min}$\\
\rule{0mm}{6mm} 3 & $\mathcal{G}_3$ & $L_8$   & $C_3 = (8,2,4)$ & $8\delta_{\min}$\\
\rule[-2mm]{0mm}{8mm} 4 & $\mathcal{G}_4=2\mathcal{G}$  & $2\mathbb{Z}^8$  & $C_4 = (8,0,\infty)$
& $16\delta_{\min}$\\\hline
\end{tabular}
\end{center}
\caption{The Golden code partition chain with corresponding
lattices, binary codes, and minimum  determinants.}
\label{tablechain}
\end{table}

\subsection{Golden subcodes}
Let us consider a subcode $\mathcal{G}_1$ obtained as right
principal ideal of the Golden code $\mathcal{G}$ \cite{David05}. In
particular we consider the subcode $\mathcal{G}_1 = \{ X B, X \in
\mathcal{G}\}$, where
\begin{equation}
B=\left[
\begin{array}{cc}
i(1-\theta) & 1-\theta \\
i\theta & i\theta %
\end{array}%
\right]. \label{Bmatrix}
\end{equation}
Since $B$ has the determinant of $1+i$, the minimum determinant of
$\mathcal{G}_1$ will be $2\delta_{\min}$.

Similarly, we consider the subcodes $\mathcal{G}_k
\subseteq\mathcal{G}$ for $k=1,\ldots, 4$, defined as
\begin{equation}
\mathcal{G}_k = \{ X B^{k}, X \in \mathcal{G}\}, \label{subcode}
\end{equation}
which provide the minimum determinant $2^k\delta_{\min}$ (see Table
\ref{tablechain}).

In the previous section we have seen how the Golden codewords
correspond to the rotated  $\mathbb{Z}^8$ lattice points. Neglecting
the rotation matrix $\mathbf{R}$, we can define an isomorphism
between $\mathcal{G}$ and $\mathbb{Z}^8$.   All the subcodes of
$\mathcal{G}$ correspond to particular sublattices of $\mathbb{Z}^8$
which are listed in Table \ref{tablechain}. In particular, it can be
shown that the codewords of $\mathcal{G}_2$, when vectorized,
correspond to Gosset lattice points $E_8$ (see Appendix I).
Similarly, we find that $\mathcal{G}_1$ corresponds to the lattice
$D_4^2$ (the direct sum of two four-dimensional Sh\"afli lattices)
and $\mathcal{G}_3$ corresponds to an eight-dimensional lattice that
is denoted by $L_8$. Finally, since $B^4=2\mathbf{I}_2$, we get the
scaled Golden code $2\mathcal{G}$ corresponding to $2\mathbb{Z}^8$.

Appendix II provides a simple overview of two basic techniques,
which will play a key role in rest of the paper: {\em Construction
A} for lattices \cite{Spherepacking} and {\em lattice set
partitioning} by coset codes \cite{Forney1,Forney2}.

As described in Appendix II, since the subcodes of $\mathcal{G}$ are
nested, the corresponding lattices form the following {\em lattice
partition chain}
\begin{equation} \label{latticechain}
\mathbb{Z}^8 \supset D_4^2 \supset E_8 \supset L_8 \supset
2\mathbb{Z}^8 .
\end{equation}
Any two consecutive lattices $ \Lambda_k \supset \Lambda_{k+1} $ in
this chain forms a four way partition, i.e., the quotient group
$\Lambda_k/\Lambda_{k+1}$ has order 4. Let
$[\Lambda_k/\Lambda_{k+1}]$ denote the set of coset leaders of the
quotient group $\Lambda_k/\Lambda_{k+1}$.

The lattices in the partition chain can be obtained by Construction
A, using the nested sequence of linear binary codes $C_k$ listed in
Table \ref{tablechain}, where $C_0$ is the universe code, $C_2$ is
the extended Hamming code or Reed-Muller code RM(1,3), $C_3$ is a
subcode of $C_2$, $C_1$ is the dual of $C_3$ and $C_4$ is the empty
code with only the all-zero codeword,  \cite{MacWilly}. The
generator matrix $G_k$ of the code $C_k$  are given by
\[G_1 = \left[
\begin{array}{cccccccc}
1 & 0 & 0 & 1 & 0 & 0 & 0 & 0  \\
0 & 1 & 0 & 1 & 0 & 0 & 0 & 0  \\
0 & 0 & 1 & 1 & 0 & 0 & 0 & 0  \\
0 & 0 & 0 & 0 & 1 & 0 & 0 & 1  \\
0 & 0 & 0 & 0 & 0 & 1 & 0 & 1  \\
0 & 0 & 0 & 0 & 0 & 0 & 1 & 1
\end{array}%
\right]\]
\[G_2 = \left[
\begin{array}{cccccccc}
0 & 1 & 0 & 1 & 0 & 1 & 0 & 1 \\
0 & 0 & 1 & 1 & 0 & 0 & 1 & 1 \\
0 & 0 & 0 & 0 & 1 & 1 & 1 & 1 \\
1 & 1 & 1 & 1 & 1 & 1 & 1 & 1
\end{array}%
\right]\]
\[G_3 = \left[
\begin{array}{cccccccc}
0 & 0 & 0 & 0 & 1 & 1 & 1 & 1 \\
1 & 1 & 1 & 1 & 1 & 1 & 1 & 1
\end{array}%
\right]\]

Looking at $G_1$ we can see that $C_1$ is the direct sum of two
parity check codes (4,3,2), this proves why it yields the lattice
$D_4^2$ by using Construction A. Similarly, since $C_3$ is the
direct sum of two repetition codes (4,1,4), we can get some insight
about the structure of the lattice $L_8$.

Following the track of \cite{Ungerboeck,Forney1,Forney2}, we
consider a partition tree of the Golden code of depth $\ell$. From a
nested subcode sequence $\mathcal{G}\supseteq \mathcal{G}_{\ell_0}
\supset \mathcal{G}_{\ell_0+1} \supset \cdots \supset
\mathcal{G}_{\ell_0+\ell}$, we have the corresponding lattice
partition chain $\mathbb{Z}^8 \supseteq \Lambda_{\ell_0} \supset
\Lambda_{\ell_0+1} \supset\cdots \supset \Lambda_{\ell_0+\ell}$
where
\begin{eqnarray*}
\Lambda_{\ell_0} &=& \Lambda_{\ell_0+1} +
[\Lambda_{\ell_0}/\Lambda_{\ell_0+1}] = \cdots \\
&=& \Lambda_{\ell_0+\ell} + [\Lambda_{\ell_0}/\Lambda_{\ell_0+1}] +
\cdots + [\Lambda_{\ell_0+\ell-1}/\Lambda_{\ell_0+\ell}]\\
&=& \Lambda_{\ell_0+\ell} + [C_{\ell_0}/C_{\ell_0+1}] + \cdots +
[C_{\ell_0+\ell-1}/C_{\ell_0+\ell}]
\end{eqnarray*}
This results in four way {\em partition tree} of depth $\ell$.
Fig.~\ref{Fig:setpartition} shows an example for $\ell=2$.

The coset leaders in $[C_k/C_{k+1}]$ form a group of order 4
isomorphic to the group $\mathbb{Z}/2\mathbb{Z} \times
\mathbb{Z}/2\mathbb{Z}$, which is generated by two binary generating
vectors ${\bf h}_1$ and ${\bf h}_2$, i.e.,
\[
[C_k/C_{k+1}] = \left\{ b_1 {\bf h}_1 + b_2 {\bf h}_2~|~b_1,b_2\in
GF(2) \right\}
\]
If we consider all the lattices in (\ref{latticechain}) and the
corresponding sequence of nested codes $C_k$, we have the following
quotient codes:
\begin{eqnarray} \label{eq:bincosetlead}
[C_0/C_1]&:& \left\{\begin{array}{l} {\bf h}^{(0)}_1 = (0,0,0,0,0,0,0,1)\\
  {\bf h}^{(0)}_2 = (0,0,0,1,0,0,0,0)\end{array} \right. \\\nonumber
[C_1/C_2] &:& \left\{\begin{array}{l} {\bf h}^{(1)}_1=(0,0,0,0,0,1,0,1)\\
  {\bf h}^{(1)}_2 = (0,0,0,0,0,0,1,1)\end{array} \right. \\\nonumber
[C_2/C_3]&:& \left\{\begin{array}{l} {\bf h}^{(2)}_1 = (0,1,0,1,0,1,0,1)\\
  {\bf h}^{(2)}_2 = (0,0,1,1,0,0,1,1)\end{array} \right. \\\nonumber
[C_3/C_4]&:& \left\{\begin{array}{l} {\bf h}^{(3)}_1=(0,0,0,0,1,1,1,1)\\
  {\bf h}^{(3)}_2 = (1,1,1,1,1,1,1,1)\end{array} \right.
\end{eqnarray}
Note that in order to generate any quotient code
$[C_{\ell_0}/C_{\ell_0+\ell}]$, we stack the above vectors in the
generator matrix $H(\ell_0,\ell_0+\ell)$ defined as
\begin{equation}
H(\ell_0,\ell_0+\ell) =
\left(\begin{array}{l}{\bf h}^{(\ell_0)}_1\\ {\bf h}^{(\ell_0)}_2\\ \vdots\\
{\bf h}^{(\ell_0+\ell-1)}_1\\ {\bf h}^{(\ell_0+\ell-1)}_2
\end{array}\right),
\end{equation}
so we can write
\begin{equation} \label{cosetsCC}
[C_{\ell_0}/C_{\ell_0+\ell}] = \left\{ (b_0,b_1,\ldots,
b_{2\ell_0+2\ell-2},b_{2\ell_0+2\ell-1} )H(\ell_0,\ell_0+\ell)
~|~b_k\in {\rm GF}(2) \right\}~.
\end{equation}
For example, to generate $[C_0/C_2]$ we use the four generators to
get the 16 coset leaders as
\begin{equation} \label{cosetsC0C2}
[C_0/C_2] = \left\{ (b_0,b_1,b_2,b_3) H(0,2)~|~b_k\in {\rm GF}(2),
H(0,2) =
\left(\begin{array}{l}{\bf h}^{(0)}_1\\ {\bf h}^{(0)}_2\\
{\bf h}^{(1)}_1\\ {\bf h}^{(1)}_2\\
\end{array}\right)  \right\}~.
\end{equation}
Note that since the $C_2=(8,4,4)$ code is self-dual, i.e.,
$C_2=C_2^\perp$ \cite{MacWilly}, we have
\[
H(2,4)=\left(\begin{array}{l}{\bf h}^{(2)}_1\\ {\bf h}^{(2)}_2\\
{\bf h}^{(3)}_1\\ {\bf h}^{(3)}_2\\
\end{array}\right) = G_2 ~.
\]

\subsection{Encoding and decoding the Golden subcodes}
In this section, we first show how to carve a cubic shaped finite
constellation from the infinite lattices corresponding to the Golden
subcodes. Construction A (Appendix II) is the design tool that also
simplifies bit labeling for such a finite constellation. We then
discuss the relation between rate and average energy required to
transmit the constellation points. Finally, we analyze the decoding
of the finite constellation.


We consider the sublattice $\Lambda_k\subseteq \mathbb{Z}^8$ at
level $k$ in the partition chain and the eight-dimensional bounding
region ${\cal B} = {\cal B}_{\rm QAM}^4$, the four-fold Cartesian
product of the bounding region of the $Q$--QAM symbols. For example,
using square QAM constellations, we have an eight-dimensional
hypercube as bounding region.

Using Construction A, a constellation point $\mathbf{x} \in
\Lambda_k \cap {\cal B}$ can be written as
\begin{equation}\label{Lambda_k_point}
\mathbf{x} = 2 \mathbf{u} + \mathbf{c}
\end{equation}
where $\mathbf{u}=(u_0,\ldots ,u_7)$ is a $8$-dimensional vector
with integer components and $\mathbf{c}=(c_0,\ldots ,c_7)$ is a
binary codeword of the corresponding code $C_k$. With an abuse of
notation we have lifted the binary components $c_i\in GF(2)$ to
integers.

Each pair of components $(2u_{2i},2u_{2i+1})$ is in ${\cal B}_{\rm
QAM}$, $i=0,1,2,3$. Note that there are only $Q/4 = 2^{\eta-2}$
distinct points from the $Q$--QAM that correspond to pairs of
components $(2u_{2i},2u_{2i+1}) \in {\cal B}_{\rm QAM}$. Since the
components $c_i$ are either 0 or 1, we are guaranteed that
$(x_{2i},x_{2i+1}) \in {\cal B}_{\rm QAM}$ and $\mathbf{x} \in {\cal
B}$.

We are now able to define the bit labels for the finite
constellation as follows. We use $q_2=8-2k$ bits to label the
$2^{q_2}$ codewords of $C_k$, through the generator matrix $G_k$,
and $q_3 = 4(\eta-2)$ bits to label the $2\mathbf{u}\in {\cal B}$.

As an example, the $E_8$ encoder structure is shown in
Fig.~\ref{Fig:E8encoder}. Assuming 16--QAM symbols ($\eta=4$), we
use $q_3=8$ bits to label the $2\mathbb{Z}^8\cap {\cal B}$ points
and $q_2=4$ bits to select one of the codewords of $C_2$ as
\begin{equation}
\mathbf{c} = ( b_1, b_2, b_3, b_4)\, G_2~. \label{cosetleader}
\end{equation}
Note that there are 16 possible codewords of $C_2$.

We observe that the constellation $\Lambda_k \cap {\cal B}$ requires
higher energy to transmit the same number of bits as the {\em
uncoded Golden code constellation} $\mathbb{Z}^8 \cap {\cal B}'$,
since ${\cal B}'\subset {\cal B}$. In particular we have that
vol(${\cal B}'$)/vol(${\cal B}$)=$N_c$ the index of the sublattice
$\Lambda_k$ over $2\mathbb{Z}^8$.

For example, encoding 12 bits with $E_8$  requires the average
energy of the 16-QAM ($E_{s,1}=2.5$), while encoding the same number
of bits with the uncoded Golden code only requires the average
energy of an 8-QAM ($E_{s,2}=1.5$). Similarly, using 128-QAM
($E_{s,1}=20.5$) we encode 24 bits with the $E_8$ lattice
constellation, while with an uncoded Golden code constellation we
can use 64-QAM with half the energy requirements ($E_{s,2}=10.5$).

Let us consider the decoding problem for $\Lambda_k\cap {\cal B}$
finite constellation. Sphere decoding of finite constellations
requires high additional complexity to handle the boundary control
problem, when the constellation does not have a cubic shape
\cite{Viterbo99}. In order to avoid this problem we adopt the
following strategy.

Given the received point $\mathbf{y}$, the lattice decoder first
minimizes the $N_c = |\Lambda_k/2\mathbb{Z}^8|$ squared Euclidean
distances in each coset
\begin{equation}
d_{j}^{2} = \min_{\mathbf{u}^{(j)} \in {\mathbb{Z}^8}}\left\Vert
\tilde{\mathbf{y}}^{(j)} - 2{\cal H}\mathbf{R}\mathbf{u}^{(j)}
\right\Vert ^{2}, ~~~~~~~ j=1,\ldots, N_c \label{latticedecodingE8}
\end{equation}
where $\tilde{\mathbf{y}}^{(j)} = \mathbf{y} - {\cal
H}\mathbf{R}{\mathbf{c}^{(j)}}, j=1,\ldots, N_c$, then makes the
final decision as
\begin{equation}
\hat{\mathbf{u}} = \mbox{arg}\, \min_j \left( d_{j}^{2}\right)~.
\end{equation}
Even if we perform $N_c$ sphere decoding operations, this strategy
is rather efficient, since each decoder is working on
$2\mathbb{Z}^8$ and visits on average an extremely low number of
lattice points during the search. In fact, this is equivalent to
working on the lattice $\mathbb{Z}^8$ at a much higher
signal-to-noise ratio.

\subsection{Performance of the Golden subcodes}
In order to compensate for the rate loss of any subcode, a
constellation expansion is required, as noted in the previous
section. For large QAM constellations, it can be seen that energy
increases approximately by a factor of $\sqrt{2}$ (1.5dB) from one
partition level to the next. Since the minimum  determinant doubles
at each partition level, we conclude that the asymptotic coding gain
(\ref{eq:asymcodgain}) is 1 (0dB). However, for small
constellations, the energy does not double and some gain still
appears.

To illustrate the observations, we show the performance of
$\mathcal{G}$ and $\mathcal{G}_2$ in Figs.~\ref{Fig:PerfE8Gold128}
and \ref{Fig:PerfE8Gold16}, corresponding to different spectral
efficiencies. In Fig.~\ref{Fig:PerfE8Gold128}, we show the
performance of $\mathcal{G}$ with 64--QAM symbols ($4\times 6 =24$
bits per codeword) and $\mathcal{G}_2$ with 128--QAM symbols
($4\times (7-2)+4=24$ bits per codeword), corresponding to a
spectral efficiency of 12 bpcu. We can see that both codes have
approximately the same codeword error rate (CER). This agrees with
the expected asymptotic coding gain
\[
\gamma_{as} =
\frac{\sqrt{4\delta_{\min}}/20.5}{\sqrt{\delta_{\min}}/10.5} = 1.02~
\rightarrow ~ 0.1\;\mbox{dB.}
\]
Fig.~\ref{Fig:PerfE8Gold16} compares the performance of the
$\mathcal{G}$ with 8--QAM symbols ($4\times 3 =12$ bits per
codeword) and $\mathcal{G}_2$ with 16--QAM symbols
($4\times(4-2)+4=12$ bits per codeword), corresponding to the
spectral efficiency of 6 bpcu. We can see that the $\mathcal{G}_2$
outperforms $\mathcal{G}$ by 0.7dB at CER of $10^{-3}$, in line
with the expected asymptotic coding gain%
\[
\gamma_{as} =
\frac{\sqrt{4\delta_{\min}}/2.5}{\sqrt{\delta_{\min}}/1.5}=
1.2~\rightarrow ~0.8\;\mbox{dB.}
\]
This small gap is essentially due to the higher energy of the
8--QAM\footnote{This is the Cartesian product of a 4--PAM and 2
2--PAM constellation.}, for which $E_{s,2}=1.5
> 2.5/\sqrt{2}$.

It is interesting to note that the $E_8$ constellation is the
densest sphere packing in dimension 8. This implies that
$\mathcal{G}_2$ maximizes
\[
\min_{X\in\mathcal{G}_2,X \neq 0} \mbox{Tr}\;(X X^\dagger) =
\min_{X\in\mathcal{G}_2,X \neq 0} \|X\|_F^2
\]
among all subcodes of the Golden code. Code design based on this
parameter is known as a trace or Euclidean distance design
criterion \cite[Sec.~10.9.3]{Biglieri05}. Our result shows how
this design criterion becomes irrelevant even at low SNR, when
using the Golden code as a starting point.

\section{Trellis Coded Modulation}\label{sec:TCM}
In this section we show how a trellis code can be used as an outer
code encoding across the Golden code inner symbols $X_{t},
t=1,\ldots,L$. We analyze the systematic design problem  of this
concatenated scheme by using Ungerboeck style set partitioning rules
for coset codes \cite{Ungerboeck,Forney1,Forney2}. The design
criterion for the trellis code is developed in order to maximize
$\Delta'_{\min}$, since this results in the maximum lower bound on
the asymptotic coding gain of the GST-TCM over the uncoded system
\begin{equation}
\gamma_{as} \geq
\frac{\sqrt{\Delta'_{\min}}/E_{s,1}}{\sqrt{\delta_{\min}}/E_{s,2}} =
\gamma'_{as}.
\end{equation}
We note that the asymptotic coding gain gives only a rough estimate
of the actual coding gain. Nevertheless, it is currently the only
means to obtain a tractable design rule for space-time TCM schemes
\cite{Tarokh98}. We then show several examples of the above schemes
with different rates and decoding complexity. We compare the
performance of such schemes with the uncoded Golden code case.

\subsection{Design criteria for GST-TCM}

{\bf Encoder structure} -- In a standard TCM encoder the trellis
encoder output is used to label the signal subset, while the uncoded
bits select the signals within the subset and yield the so called
parallel transitions in the trellis \cite{Biglieri05}.
Fig.~\ref{Fig:systemmodel} shows the encoder structure of the
proposed concatenated scheme. The input bits feed two encoders, an
upper {\em trellis encoder} and a lower {\em sublattice encoder}.
The output of the trellis encoder is used to select the coset, while
the sublattice encoder will select the point within the coset. The
trellis will have parallel transitions on each branch corresponding
to the constellation points within the same coset.

We consider two lattices $\Lambda_{\ell_0}$ and
$\Lambda_{\ell_0+\ell}$ from the lattice partition chain in Table
\ref{tablechain}, such that $\Lambda_{\ell_0+\ell}$ is a proper
sublattice of the lattice $\Lambda_{\ell_0}$, where $\ell$ denotes
the {\it relative partition level} of $\Lambda_{\ell_0+\ell}$ with
respect to $\Lambda_{\ell_0}$. Let $\ell_{0}$ denote the {\it
absolute partition level} of the lattice $\Lambda_{\ell_0}$. For
example, with $\ell_{0} = 0, \ell = 2$, we have $\Lambda_{\ell_0} =
\mathbb{Z}^{8}$ and $\Lambda_{\ell_0+\ell} = E_8$, with $\ell_{0} =
2, \ell = 2$, we have $\Lambda_{\ell_0} = E_8$ and
$\Lambda_{\ell_0+\ell} = 2\mathbb{Z}^{8}$.

The quotient group $\Lambda_{\ell_0}/\Lambda_{\ell_0+\ell}$ has
order
\begin{equation}
N_c = \left\vert \Lambda_{\ell_0}/\Lambda_{\ell_0+\ell} \right\vert
= 4^\ell,
\end{equation}
which corresponds to the total number of cosets of the sublattice
$\Lambda_{\ell_0+\ell}$ in the lattice $\Lambda_{\ell_0}$.

Let us consider a trellis encoder operating on $q_1$ information
bits. Given the relative partition depth $\ell$, we need to select
$N_c=2^{2\ell}$ distinct cosets. If we consider a trellis code with
rate $R_c = 1/\ell$, the trellis encoder must output
\[
n_c = q_1/R_c = 2\ell =\log_2(N_c) \mbox{~~~bits,}
\]
hence we can input $q_1=2$ bits. Since the trellis has $2^{q_1}$
incoming and outgoing branches from each state, this choice is made
to preserve a reasonable trellis branch complexity. The previous
design, proposed in \cite{David05}, had a much larger branch
complexity.

The $n_c$ bits are used by the coset mapper to label the coset
leader $\mathbf{c}_1 \in [C_{\ell_0}/C_{\ell_0+\ell}]\sim
[\Lambda_{\ell_0}/\Lambda_{\ell_0+\ell}]$. The mapping is obtained
by the product of the $n_c$ bit vector with a binary  coset leader
generator matrix
\begin{equation} \label{eq:cosetgenmat}
H_{c_{1}} = \left(\begin{array}{l}{\bf h}^{(\ell_0)}_1\\ {\bf h}^{(\ell_0)}_2\\ \vdots\\
{\bf h}^{(\ell_0+\ell-1)}_1\\ {\bf h}^{(\ell_0+\ell-1)}_2
\end{array}\right),
\end{equation}
where the rows are taken from (\ref{eq:bincosetlead}).

We assume that we have a total of $4q=q_1+q_2+q_3$ input information
bits. The lower encoder is a sublattice encoder for
$\Lambda_{\ell_0+\ell}$ and operates on the remaining $q_2 + q_3$
information bits.  The $q_2 = 2\times (4-\ell - \ell_{0})$ bits
label the cosets of $2\mathbb{Z}^{8}$ in $\Lambda_{\ell_0+\ell}$ by
multiplying the following binary generator matrix
\begin{equation} \label{eq:sublattice}
H_{c_{2}} = \left(\begin{array}{l}{\bf h}^{(\ell_0+\ell)}_1\\ {\bf h}^{(\ell_0+\ell)}_2\\ \vdots\\
{\bf h}^{(3)}_1\\ {\bf h}^{(3)}_2
\end{array}\right),
\end{equation}
which generates coset leader $\mathbf{c}_2 \in
[\Lambda_{\ell_0+\ell}/{2\mathbb{Z}^{8}}]$. We finally add both
coset leaders of $\mathbf{c}_1$ and $\mathbf{c}_2$ modulo 2 to get
$\mathbf c'$. The remaining $q_3 = 4q - q_1 - q_2$ bits go through
$2\mathbb{Z}^8$ encoder and generate vector $\mathbf{2u}$ as
detailed in Appendix II. Finally, $\mathbf{2u}$ is added to
$\mathbf{c'}$ (lifted to have integer components) and mapped to the
Golden codeword $X_t$.

We now focus on the structure of the trellis code to be used. We consider
linear convolutional encoders over the quaternary alphabet
$\mathbb{Z}_4=\{0,1,2,3\}$ with mod 4 operations. We assume the natural
mapping between pairs of bits and $\mathbb{Z}_4$ symbols, i.e., $0\rightarrow
00,1\rightarrow 01, 2 \rightarrow 10, 3 \rightarrow 11$. Let $\beta\in \mathbb{Z}_4$
denote the input symbol and $\alpha_1,\ldots,\alpha_{\ell}\in \mathbb{Z}_4$
denote the $\ell$ output symbols generated by the generator polynomials
$g_1(D), \ldots g_\ell(D)$ over $\mathbb{Z}_4$.

For example, Figure \ref{Fig:trellis4states} shows a 4 state encoder
with rate $R_c=1/2$ defined by the generator polynomials $g_1(D)=1$
and $g_2(D)=D$. The trellis labels for outgoing and incoming
branches listed from top to bottom. Figure \ref{Fig:setpartition}
shows how the $N_c=16$ cosets can be addressed through a partition
tree of depth 2.

{\bf Labeling} -- Let us first consider the conventional design of
the trellis labeling in a TCM scheme. We then show how this can be
directly transferred to GST-TCM. The conventional TCM design
criteria attempt to increase the minimum Euclidean distance
$d_{\min}$ between codewords in the following way.
\begin{enumerate}
\item
Use subconstellations with a larger minimum Euclidean distance
$d_{p,\min}$, known as {\em intra-coset distance}
\item  Label the parallel branches in the trellis with the points
within the same subconstellation.
\item
Label the trellis branches for different states so that the
partitions can increase the {\em inter-coset distance} $d_{s,\min}$
among code sequences.
\end{enumerate}

The aim of our GST-TCM design criteria is to maximize the lower
bound $\Delta_{\min}'$ in (\ref{lowerbound}). The additive structure
of the $\Delta_{\min}'$ enables to use the same strategy that is
used for the Euclidean distance in conventional TCM design. Let
\begin{equation}
\Delta_{\rm p}=2^{\ell_0+\ell}\delta_{\min}
\end{equation}
denote the minimum determinant on the trellis parallel transitions
corresponding to the Golden code partition $\Lambda_{\ell_0+\ell}$
of absolute level $\ell_0+\ell$. Let
\begin{equation}
\Delta_{\rm s}=\min_{\mathbf{X}\neq{\bf 0}_{2\times 2L }}
\sum_{t=t_o}^{t_o+L'-1} \det(X_tX_t^\dagger) \label{eq:Delta_simple}
\end{equation}
denote the minimum determinant on the shortest simple error event,
where $L'$ is the length of the shortest simple error event
diverging from the zero state at $t_o$ and merging to the zero state
at  $t_i = t_o+L'$. We can increase $\Delta_{\rm s}$ in
(\ref{eq:Delta_simple}) either by increasing $L'$ or by increasing
the $\det(X_tX_t^\dagger)$ terms. Fig. \ref{Fig:intercoset} shows
the possible inter coset distances contributing to
(\ref{eq:Delta_simple}).

Note that once $L'$ is fixed, Ungerboeck's design rules focus on the
first and last term only. The lower bound $\Delta_{\min}'$ in
(\ref{lowerbound}) is determined either by the parallel transition
error events or by the shortest simple error events in the trellis,
i.e.,
\begin{equation} \label{eq:simplepath}
\Delta_{\min}' = \min\left\{\Delta_{\rm p}, \Delta_{\rm s} \right\}
\geq \min\left\{\Delta_{\rm p},
\min_{X_{t_o}}\det(X_{t_o}X_{t_o}^\dagger) +
\min_{X_{t_i}}\det(X_{t_i}X_{t_i}^\dagger) \right\}.
\end{equation}
The corresponding coding gain will be
\begin{equation} \label{eq:simplepathgains}
\gamma'_{as} = \min\left\{\gamma'_{as}(\Delta_{\rm p}),
\gamma'_{as}(\Delta_{\rm s}) \right\} .
\end{equation}
Therefore, we can state the following:\\
 {\bf Design Criterion} -- {\em
We focus on $\Delta_{\min}'$. The incoming and outgoing branches for
each state should belong to different cosets that have the common
father node as deep as possible in the partition tree. This
guarantees that simple error events in the trellis give the largest
contribution to $\Delta_{\min}'$.}

In order to fully satisfy the above criterion for a given relative
partition level $\ell$, the minimum number of trellis states should
be $N_c=2^{2\ell}$. In order to reduce complexity we will also
consider trellis codes with fewer states. We will see in the
following that the performance loss of these suboptimal codes (in
terms of the above design rule) is marginal since $\Delta_{\rm p}$
is dominating in (\ref{eq:simplepath}). Nevertheless, the
optimization of $\Delta_{\rm s}$ yields a performance enhancement.
In fact, maximizing $\Delta_{\rm s}$ has the effect of minimizing
another relevant PWEP term.

{\bf Decoding} -- Let us analyze the decoding complexity. The
decoder is structured as a typical TCM decoder, i.e. a Viterbi
algorithm using a branch metric computer. The branch metric computer
should output the distance of the received symbol from all the
cosets of $\Lambda_{\ell_0+\ell}$ in $\Lambda_{\ell_0}$.  The
decoding complexity depends on two parameters
\begin{itemize}
\item $N_c$ the total number of distinct parallel branch metrics
\item the number of states  in the trellis.
\end{itemize}
We observe that the branch metric computer can be realized either as a
traditional sphere decoder for each branch or as single list sphere decoder
which can keep track of all the cosets at once.

\subsection{Code Design Examples for TCM}
In this subsection, we give four examples of GST-TCM with different
numbers of states using different partitions
$\Lambda_{\ell_0}/\Lambda_{\ell_0+\ell}$. We assume a frame length
$L=130$ in all examples. All related parameters are summarized in
Table \ref{table2}.

The trellis code generator polynomials have been selected by an
exhaustive search among all polynomials of degree less than four
with quaternary coefficients. The selection was made in order to
satisfy the design criterion (when possible) and to maximize
$\Delta_{s,\min}$.

We first describe the uncoded Golden code schemes, which are used as
reference systems for performance comparison. In the standard
uncoded Golden code, four $Q$--QAM information symbols are sent for
each codeword (\ref{goldencodeword}), for a total of $4q$
information bits, where $q=\log_2(Q)$. When $q$ is not integer, we
have to consider different size QAM symbols within the same Golden
codeword, as shown in the following examples.
\begin{itemize}
\item
{\bf 5bpcu --} A total of 10 bits must be sent in a Golden codeword:
the symbols $a$ and $c$ are in a 4-QAM (2bits), while  the symbols
$b$ and $d$ are in a 8-QAM (3bits). This guarantees that the same
average energy is transmitted from both antennas. In this case we
have $E_{s,2}=(0.5+1.5)/2= 1$ and $q=2.5$ bits.
\item
{\bf 6bpcu --} A total of 12 bits must be sent in a Golden codeword:
the symbols $a,b,c,d$ are in a 8-QAM (3bits). In this case we have
$E_{s,2}=1.5$ and $q=3$ bits.
\item
{\bf 7bpcu --} A total of 14 bits must be sent in a Golden codeword:
the symbols $a$ and $c$ are in a 8-QAM (3bits), while the symbols
$b$ and $d$ are in a 16-QAM (4bits). This guarantees that the same
average energy is transmitted from both antennas. In this case we
have $E_{s,2}=(1.5+2.5)/2= 2$ and $q=3.5$ bits.
\item
{\bf 10bpcu --} A total of 20 bits must be sent in a Golden
codeword: the symbols $a,b,c,d$ are in a 32-QAM (5bits). In this
case we have $E_{s,2}=5$ and $q=5$ bits.
\end{itemize}

{\bf Example 1 --} {\it We use a two level partition
$E_8/2\mathbb{Z}^8$. The 4 and 16 state trellis codes using 16--QAM
($E_{s,1}=2.5$) gain 2.2dB and 2.5dB, respectively, over the uncoded
Golden code ($E_{s,2}=1$) at the rate of 5bpcu.}

The two level partition ($\ell_0 = 2$ and $\ell = 2$) has a quotient
group $E_8/ 2\mathbb{Z}^8$  of order $N_{c}=16$. The quaternary
trellis encoders for 4 and 16 states with rate $R_{c} = 1/2$, have
$q_1=2$ input information bits and $n_c=4$ output bits, which label
the coset leaders using the generator matrix with rows ${\bf
h}^{(2)}_1, {\bf h}^{(2)}_2, {\bf h}^{(3)}_1, {\bf h}^{(3)}_2$.  The
trellis structures are shown in Fig.~\ref{Fig:trellis4states} and
Fig.~\ref{Fig:trellis16states}, respectively. The sublattice encoder
has $q_2 = 0$ and  $q_3 = 8$ input bits, giving a total number of
input bits per information symbol $q=(q_1+q_2+q_3)/4=10/4=2.5$bits.

In Fig.~\ref{Fig:trellis4states}, for each trellis state, the four
outgoing branches with labels $\alpha_{1}, \alpha_{2}$,
corresponding to input $\beta=0,1,2,3$, are listed on the left side
of the trellis. Similarly, four incoming trellis branches to each
state are listed on the right side of the trellis structure. In this
case, $\alpha_{1}$ chooses the cosets from $L_{8}$ in $\Lambda=E_8$
and $\alpha_{2}$ chooses the cosets from $\Lambda_\ell =
2\mathbb{Z}^{8}$ in $L_{8}$.

We can observe that the four branches merging in each state belong
to four different cosets of 2$\mathbb{Z}^8$ in ${L}_8$, since
$\alpha_1$ is constant and $\alpha_2$ varies (see
Fig.~\ref{Fig:setpartition}). This guarantees an increased
$\Delta_{\min}'$. On the other hand, the four branches departing
from each state are in the cosets of ${L}_8$ in ${E}_8$. This does
not give the largest possible $\Delta_{\min}'$ since $\alpha_1$
varies. Looking for example at the zero state, there are four
outgoing branches labeled by $\alpha_1 = 0,1,2,3$ and $\alpha_2$ is
fixed to $0$, while the four incoming branches are labeled by
$\alpha_1 = 0$ and $\alpha_2 = 0,1,2,3$.

This results in a suboptimal design since it can not guarantee that
the outgoing trellis paths belong to cosets that are in the deepest
level ($2\mathbb{Z}^{8}$) of the partition tree. We can see that the
shortest simple error event has a length of $L'=2$, corresponding to
the state sequence $0\rightarrow 1\rightarrow 0$ and labels $10,01$.
This yields the lower bound on the asymptotic coding gain
\begin{equation}
\gamma'_{as} = \frac{\sqrt{\min(16\delta_{\min},
4\delta_{\min}+8\delta_{\min})}/E_{s,1}}{\sqrt{\delta_{\min}}/E_{s,2}}
 ~\rightarrow~  1.4 ~\mbox{dB}.
\end{equation}

The above problem suggests the use of a 16 state encoder. In
Fig.~\ref{Fig:trellis16states}, we can see that the shortest simple
error event has length $L'= 3$  corresponding to the state sequence
$0\rightarrow 1\rightarrow 4\rightarrow 0$ and labels $01,10,01$. In
general, we have that the first output label $\alpha_1$ is fixed for
both outgoing and incoming states. This guarantees both incoming and
outgoing trellis branches from each state belong cosets with the
deepest father nodes in the partition tree. This yields the lower
bound on the corresponding asymptotic coding gain
\begin{equation}
\gamma'_{as} = \frac{\sqrt{\min(16\delta_{\min},
8\delta_{\min}+4\delta_{\min}+8\delta_{\min})}/E_{s,1}}{\sqrt{\delta_{\min}}/E_{s,2}}
  ~\rightarrow~ 2.0 ~\mbox{dB}.
\end{equation}

Compared to 4 state, the 16 state GST-TCM has a higher decoding
complexity. It requires $64$ lattice decoding operations in each
trellis section, while the 4 state GST-TCM only requires $16$
lattice decoding operations. Note that each lattice decoding
operation is working on $2\mathbb{Z}$.

Performance comparison of the proposed codes with the uncoded scheme
with 5 bpcu is shown in Fig.~\ref{Fig:8}. We can observe that a
simple 4 state GST-TCM outperforms the uncoded scheme by 2.2dB at
the FER of $10^{-3}$. The 16-state GST-TCM outperforms the uncoded
case by 2.5dB at the FER of $10^{-3}$.
\begin{table}[t]
\begin{center}
\scriptsize
\begin{tabular}{|c||c|c|c|c|c|c|c|c|c|c|c|c|c|c|} \hline
\rule{-1mm}{5mm} Ex. & $\Lambda$ & $\Lambda_\ell $ & $\ell_0$ &
$\ell$ & $q_1$& $q_2$ & $q_3$ & bpcu & $Q$ & states &
$g_1(D),\ldots , g_\ell(D) $ & $\begin{array}{c}
  \gamma'_{as} \\
  (\Delta_{\rm p}) \\
\end{array}$&$\begin{array}{c}
  \gamma'_{as} \\
  (\Delta_{\rm s}) \\
\end{array}$& $\begin{array}{c}
  \text{gain} \\
  \text{@} \\
  10^{-3} \\
\end{array}$\\
\hline\hline \rule{0mm}{5mm} $1$   & $E_8$  & $2\mathbb{Z}^8$  &
$2$
& $2$ & $2$ & $0$ & $8$ & $5$ & $16$ &  $4$ & $\left ( 1, D \right )$ & $2.0$ &$1.4$ & $2.2$ \\
\rule{-1mm}{5mm} &&&&&&&&&& $16$ & $\left ( D, 1 + D^2 \right )$ & $2.0$ &$2.5$& $2.5$\\
\hline \rule{0mm}{5mm} $2$   & $\mathbb{Z}^8$  & $E_8$  & $0$ &
$2$
& $2$ & $4$ & $8$ & $7$ & $16$ & $4$ & $\left ( 1, D \right )$ & $2.0$ &$1.4$& $3.0$ \\
\rule{-1mm}{5mm} &&&&&&&&&& $16$ & $\left ( D, 1 + D^2 \right )$ & $2.0$ &$2.5$& $3.3$\\
\hline \rule[-2mm]{0mm}{7mm} $3$   & $\mathbb{Z}^8$  & $L_8$  &
$0$
& $3$ & $2$ & $2$ & $8$ & $6$ & $16$ & $16$ & $\left ( D, D^2, 1 + D^2 \right )$ & $2.3$ &$2.0$& $4.2$\\
\rule{-1mm}{5mm} &&&&&&&&&& $64$ & $\left ( D, D^2, 1 + D^3 \right )$ & $2.3$ &$3.0$& $4.3$\\
\hline \rule[-2mm]{0mm}{7mm} $4$   & $\mathbb{Z}^8$  & $L_8$  &
$0$
& $3$ & $2$ & $2$ & $16$ & $10$ & $64$ & $16$ & $\left ( D, D^2, 1 + D^2 \right )$ & $1.3$ &$1.0$& $1.5$\\
\rule{-1mm}{5mm} &&&&&&&&&& $64$ & $\left ( D, D^2, 1 + D^3 \right )$ & $1.3$ &$2.0$& $1.5$\\
\hline
\end{tabular}
\end{center}
\caption{Summary of the parameters of GST-TCM Examples } \label{table2}
\end{table}

{\bf Example 2 --} {\it We use a two level partition
$\mathbb{Z}^8/E_8$  ($\ell_0 = 0$ and $\ell = 2$). The 4 and 16
state trellis codes using 16-QAM ($E_{s,1}=2.5$) gain 3.0dB and
3.3dB, respectively, over uncoded Golden code ($E_{s,2}=2$) at the
rate of 7 bpcu}.

As in Example 1, we can see that the 4 state trellis code is
suboptimal since it can not guarantee that both the incoming and
outgoing trellis paths belong to cosets that are in the deepest
level ($E_8$) of the partition tree. In contrast, the 16 state
trellis code always has a fixed label $\alpha_1$ in each state. This
fully satisfies the proposed design criteria.  However, the 16 state
code requires higher decoding complexity. Finally, we have
\begin{equation}
\gamma'_{as} =
\frac{\sqrt{\min(4\delta_{\min},\delta_{\min}+2\delta_{\min})}
/E_{s,1}}{\sqrt{\delta_{\min}}/E_{s,2}}
 ~\rightarrow~ 1.4 ~\mbox{dB}
\end{equation}
for the 4 state GST-TCM and
\begin{equation}
\gamma'_{as} =
\frac{\sqrt{\min(4\delta_{\min},2\delta_{\min}+\delta_{\min}+2\delta_{\min})}
/E_{s,1}}{\sqrt{\delta_{\min}}/E_{s,2}}
 ~\rightarrow~ 2.0~\mbox{dB}
\end{equation}
for the 16 state GST-TCM.

Performance of both the proposed TCM and uncoded transmission (7
bpcu) schemes is compared in Fig.~\ref{Fig:9}. It is shown that the
proposed 4 and 16 state TCMs outperform the uncoded case by 3.0dB
and 3.3dB at the FER of $10^{-3}$.

Compared to Example 1, this GST-TCM has a higher decoding
complexity. It requires $N_c=256$ lattice decoding operations of
$2\mathbb{Z}^8$ in each trellis section or 16 lattice decoders of
cosets of $E_8$.

{\bf Example 3 --} {\it We use a three level partition
$\mathbb{Z}^8/L_8$ ($\ell_0 = 0$ and $\ell = 3$). The 16 and 64
state trellis codes using 16--QAM ($E_{s,1}=2.5$) gain 4.2 and 4.3
dB, respectively, over an uncoded Golden code ($E_{s,2}=1.5$) at the
rate of 6 bpcu}.

In Fig.~\ref{Fig:trellis16statesr3}, for each trellis state, the
four outgoing branches with labels $\alpha_1, \alpha_2, \alpha_3$,
corresponding to input $\beta =0,1,2,3$, are listed on the left side
of the trellis. Similarly, the four incoming trellis branches to
each state are listed on the right side of the trellis structure. In
such a case, $\alpha_1$ chooses the cosets from $D_4^2$ in
$\Lambda=\mathbb{Z}^8$, $\alpha_2$ chooses the cosets from $E_8$ in
$D_4^2$, and $\alpha_3$ chooses the cosets from $\Lambda_\ell = L_8$
in $E_8$.

The four branches departing from each state belong to four different
cosets of $L_8$, since $\alpha_1$ and $\alpha_2$ are constant, while
$\alpha_3$ varies. On the other hand, the four branches arriving in
each state are cosets of $E_8$. This does not yield the largest
possible $\Delta_{\min}'$, since only $\alpha_1$ is fixed but
$\alpha_2$ varies. This results in a suboptimal design since it can
not guarantee that both incoming and outgoing trellis paths belong
to cosets that are in the deepest level ($L_{8}$) of the partition
tree.

We can see that the shortest simple error event has a length of
$L'=3$ corresponding to the state sequence $0\rightarrow
1\rightarrow 4\rightarrow 0$ and labels $001,100,011$. This yields
the lower bound of the corresponding asymptotic coding gain
\begin{equation}
\gamma'_{as} = \frac{\sqrt{\min(8\delta_{\min},
4\delta_{\min}+\delta_{\min}+2\delta_{\min})}
/E_{s,1}}{\sqrt{\delta_{\min}}/E_{s,2}} ~\rightarrow~ 2.0~\mbox{dB}.
\end{equation}

The above problem suggests the use of a 64 state encoder. In
Fig.~\ref{Fig:trellis16states}, we can see that the shortest simple
error event has length $L'= 4$ corresponding to the state sequence
$0\rightarrow 1\rightarrow 4\rightarrow 16\rightarrow 0$ and labels
$001,100,010,001$. Note that now the output labels $\alpha_1,
\alpha_2$ are fixed for all outgoing and incoming states. This
guarantees both incoming and outgoing trellis branches from each
state belong to the cosets that are deepest in the partition tree.
This yields the lower bound of the corresponding asymptotic coding
gain
\begin{equation}
\gamma'_{as} = \frac{\sqrt{\min(8\delta_{\min},
4\delta_{\min}+\delta_{\min}+2\delta_{\min}+4\delta_{\min})}
/E_{s,1}}{\sqrt{\delta_{\min}}/E_{s,2}}  ~\rightarrow~
2.3~\mbox{dB}.
\end{equation}

Performance of the proposed codes and the uncoded scheme with 6 bpcu
is compared in Fig.~\ref{Fig:12}. We can observe that a 16 state
GST-TCM outperforms the uncoded scheme by 4.2 dB at the FER of
$10^{-3}$. The 64 state GST-TCM outperforms the uncoded case by 4.3
dB  at FER of $10^{-3}$.

Note that in this Example with 16 states, we have the same decoding
complexity as in the previous example with 16 states.

{\bf Example 4 --} {\it We use the same partition as in Example 3.
The 16 and 64 state trellis codes using 64-QAM  ($E_{s,1}=10.5$)
gain 1.5 dB, in both cases, over an uncoded Golden code
($E_{s,2}=5$) at the rate of 10}.

The trellis structures are shown in Figures
\ref{Fig:trellis16statesr3} and \ref{Fig:trellis64statesr3},
respectively. This yields the lower bounds of the corresponding
asymptotic coding gain
\begin{equation}
\gamma'_{as} =
\frac{\sqrt{\min(8\delta_{\min},4\delta_{\min}+\delta_{\min}+2\delta_{\min})}
/E_{s,1}}{\sqrt{\delta_{\min}}/E_{s,2}}  ~\rightarrow~
1.0~\mbox{dB}.
\end{equation}
for the 16 state GST-TCM and
\begin{equation}
\gamma'_{as} =
\frac{\sqrt{\min(8\delta_{\min},4\delta_{\min}+\delta_{\min}+2\delta_{\min}+4\delta_{\min})}
/E_{s,1}}{\sqrt{\delta_{\min}}/E_{s,2}}  ~\rightarrow~1.3~\mbox{dB}.
\end{equation}
for the 64 state GST-TCM.

Fig.~\ref{Fig:13} compares the performance of above codes at the
spectral efficiency of 10 bpcu with 64 QAM signal constellation for
GST-TCM and 32 QAM signal constellation for uncoded case,
respectively. It is shown that a 16 state GST-TCM outperforms the
uncoded scheme by 1.5dB at the FER of $10^{-3}$. The 64 state code
has similar performance as the 16 state code.

{\bf Remarks:}  For GST-TCM, we can see that the lower bound
$\gamma'_{as}$ on $\gamma_{as}$ is only a rough approximation of the
true system performance. This is due to the following reasons:
\begin{enumerate}
\item $\gamma_{as}$ is based on the worst case pairwise error
event which is not always the strongly dominant term of the full
union bound in fading channels; \item the lower bound
$\gamma'_{as}$ on $\gamma_{as}$ can be loose due to the
determinant inequality; \item the multiplicity of the minimum
determinant paths is not taken into account.
\end{enumerate}

Looking at Table 2, we observe that the true coding gain is better
approximated by a combination of $\gamma'_{as}(\Delta_{\rm p})$ and
$\gamma'_{as}(\Delta_{\rm s})$ in (\ref{eq:simplepathgains}), rather
than $\gamma'_{as}$.

\section{Conclusions}\label{sec:conclu}
In this paper, we presented GST-TCM, a concatenated scheme for slow
fading $2\times 2$ MIMO systems. The inner code is the Golden code
and the outer code is a trellis code. Lattice set partitioning is
designed specifically to increase the minimum determinant of the
Golden codewords, which label the branches of the outer trellis
code. Viterbi algorithm is applied in trellis decoding, where branch
metrics are computed by using a lattice sphere decoder. The general
framework for GST-TCM design and optimization is based on Ungerboeck
TCM design rules.

Simulation shows that 4 and 16 state GST-TCMs achieve 3dB and
4.2dB performance gains over uncoded Golden code at FER of
$10^{-3}$ with spectral efficiencies of 7 bpcu and 6 bpcu,
respectively.

Future work will explore the possibility of further code
optimization, by an extensive search based on the determinant
distance spectrum, which gives a more accurate approximation of the
true coding gain.

\appendix
\section*{Appendix I: Proof of (\ref{subcode})}
Let us consider a subcode $\mathcal{G}_2$ of the Golden code
$\mathcal{G}$ obtained by $\mathcal{G}_2 = \{ X B^{2}, X \in
\mathcal{G}\}$, where $B$ is given in (\ref{Bmatrix}) and $X$ is
given as
\begin{equation}
X=\left[
\begin{array}{cc}
\alpha \left( a+b\theta \right)  & \alpha \left( c+d\theta \right)  \\
i \bar{\alpha} \left( c+d\bar{\theta} \right)  & \bar{\alpha}
\left( a+b\bar{\theta} \right)
\end{array}%
\right],
\end{equation}
where we omit the normalization factor $\frac{1}{\sqrt{5}}$ for
simplicity. After manipulations, we obtain the subcode
$\mathcal{G}_2$ codeword
\begin{equation}
\left[ \begin{array}{cc}
g_{11} & g_{12} \\
g_{21} & g_{22}
\end{array}\right] = XB^{2} \label{newG}
\end{equation}
where
\begin{eqnarray}
g_{11} &=&\left[ -1-i2\left( 1+\bar{\theta}\right) \right]
a+\left( -\theta +i2\bar{\theta}\right) b+\left( -\theta +i\right)
c+\left( -1-\theta
+i\theta \right) d,  \nonumber \\
g_{21} &=&\left[ -\theta -i\left( 1+\theta \right) \right]
a+\left( 1+i\theta \right) b+\left[ \theta -i2\bar{\theta}\right]
c+\left[ -1-i\left(
2+2\bar{\theta}\right) \right] d,  \nonumber \\
g_{12}&=&\left[ -1-\bar{\theta}+i\bar{\theta}\right] a+\left( \bar{%
\theta}-i\right) b+\left( -2\theta -i\bar{\theta}\right) c+\left(
-2-2\theta
+i\right) d,  \nonumber \\
g_{22}&=&\left[ -1-i2\left( 1+\theta \right) \right] a+\left( -\bar{%
\theta}+i2\theta \right) b+\left( -1+\theta +i\right) c+\left( -1-\bar{\theta%
}+i\bar{\theta}\right) d, \nonumber
\end{eqnarray}
where $a,b,c,d \in \mathbb{Z} [i]$. Note that $\bar{\theta} = 1 -
{\theta}$ and ${\theta}^2 = \theta + 1$. Vectorizing (\ref{newG})
yields
\begin{equation}
vec\left( XB^2\right)  = \tilde{\mathbf{R}}\mathbf{u}
\end{equation}
where
\begin{equation}
vec\left( XB^2 \right)  =\left[ \Re \left( g_{11} \right),\, \Im
\left( g_{11} \right),\, \Re \left( g_{21} \right),\, \Im \left(
g_{21} \right),\, \Re \left( g_{12} \right),\, \Im \left( g_{12}
\right),\, \Re \left( g_{22} \right),\, \Im \left( g_{22} \right)
\right] ^{T}
\end{equation}
\begin{equation}
\tilde{{\mathbf{R}}} =\left[
\begin{array}{cccccccc}
-1 & 2\left( 1+\bar{\theta}\right)  & -\theta  & -2\bar{\theta} &
-\theta  &
-1 & -1-\theta  & -\theta  \\
-2\left( 1+\bar{\theta}\right)  & -1 & 2\bar{\theta} & -\theta  &
1 &
-\theta  & \theta  & -1-\theta  \\
-\theta  & 1+\theta  & 1 & -\theta  & \theta  & 2\bar{\theta} & -1 & 2+2\bar{%
\theta} \\
-1-\theta  & -\theta  & \theta  & 1 & -2\bar{\theta} & \theta  & -2-2\bar{%
\theta} & -1 \\
-1-\bar{\theta} & -\bar{\theta} & \bar{\theta} & 1 & -2\theta  &
\bar{\theta}
& -2-2\theta  & -1 \\
\bar{\theta} & -1-\bar{\theta} & -1 & \bar{\theta} & -\bar{\theta}
&
-2\theta  & 1 & -2-2\theta  \\
-1 & 2\left( 1+\theta \right)  & -\bar{\theta} & -2\theta  &
-1+\theta  & -1
& -1-\bar{\theta} & -\bar{\theta} \\
-2\left( 1+\theta \right)  & -1 & 2\theta  & -\bar{\theta} & 1 &
-1+\theta
& \bar{\theta} & -1-\bar{\theta}%
\end{array}%
\right],
\end{equation}
and
\begin{equation}
\mathbf{u} =\left[ \Re \left( a\right),\, \Im \left( a\right),\,
\Re \left( b\right),\, \Im \left( b\right),\, \Re \left(
c\right),\, \Im \left( c\right),\, \Re \left( d\right),\, \Im
\left( d\right) \right] ^{T}.
\end{equation}
The matrix $\tilde{\mathbf{R}}$ can be written as
\begin{equation}
\tilde{{\mathbf{R}}} = {\mathbf{R}}\mathbf{\tilde{M}}.  \nonumber
\end{equation}
Substituting the matrix $\mathbf{R}$, defined in
(\ref{rotationmatrix}), into above equation yields the lattice
generator matrix
\begin{eqnarray}
\mathbf{\tilde{M}} &=& {\mathbf{R}^{T} \tilde{\mathbf{R}}}= \left[
\begin{array}{cccccccc}
-2 & 1 & 1 & 0 & 0 & 0 & -1 & 0 \\
-1 & -2 & 0 & 1 & 0 & 0 & 0 & -1 \\
1 & 0 & -1 & 1 & -1 & 0 & -1 & 0 \\
0 & 1 & -1 & -1 & 0 & -1 & 0 & -1 \\
0 & -1 & 0 & 1 & -1 & 1 & -1 & 0 \\
1 & 0 & -1 & 0 & -1 & -1 & 0 & -1 \\
0 & 1 & 0 & 0 & -1 & 0 & -2 & 1 \\
-1 & 0 & 0 & 0 & 0 & -1 & -1 & -2%
\end{array}%
\right].  \nonumber
\end{eqnarray}
By conducting LLL lattice basis reduction, we found that the lattice
generator matrix $\mathbf{\tilde{M}}$ has the minimum squared
Euclidean distance $d^{2}_{min} = 4$. Since the determinant of
$\tilde{\mathbf{M}}$ is 16, the packing density coincides with the
one of $E_8$, which is the unique optimal sphere packing in 8
dimension. Note that there exist multiple lattice generator matrices
for $E_8$ lattices, all of which have the same properties as above
\cite{Spherepacking}. Therefore we conclude that the subcode
$\mathcal{G}_2$ of the Golden code $\mathcal{G}$, when vectorized,
corresponds to the $E_8$ lattice points. A similar approach can be
used for the other lattices in the partition.

\section*{Appendix II: Construction A and Set Partitioning}

In this Appendix we review the basic principles of  {\em
Construction A} and  {\em lattice set partitioning} by coset codes
following a simple example based on the lattice $\mathbb{Z}^2$. The
general theory underlying these techniques is described in detail in
\cite{Spherepacking,Forney1,Forney2}. We assume that the  reader is
familiar with the basic facts of group theory, in particular we will
use the notions of group, subgroup, quotient group, and group
isomorphism \cite{groupsbook}.

Construction A establishes a correspondence between an integer
lattice and a linear binary code \cite{Spherepacking}. In particular
given an integer lattice $\Lambda$ we obtain all the codewords of a
linear binary code $C$ by taking all components of the lattice
points mod 2, we write:
\begin{equation}
C = \Lambda \!\!\!\mod 2
\end{equation}
On the other hand given a
linear binary code $C=(n,k,d)$ with codewords ${\bf c}_i$ we can
write:
\begin{equation}
\Lambda = 2\mathbb{Z}^n + C = \bigcup_{{\bf c}_i\in C}
\left(2\mathbb{Z}^n + {\bf c}_i \right)
\end{equation}
This construction provides also a simple relation between the
minimum Hamming distance $d$ of the code and the minimum Euclidean
distance between any two lattice points. For this reason it can be
used to design dense sphere packing lattices \cite{Spherepacking}.
For our purposes we will use Construction A as means to handle the
set partitioning and to bit-label the lattice points within a finite
constellation.

\begin{figure}
\begin{center}
  \psfig{file=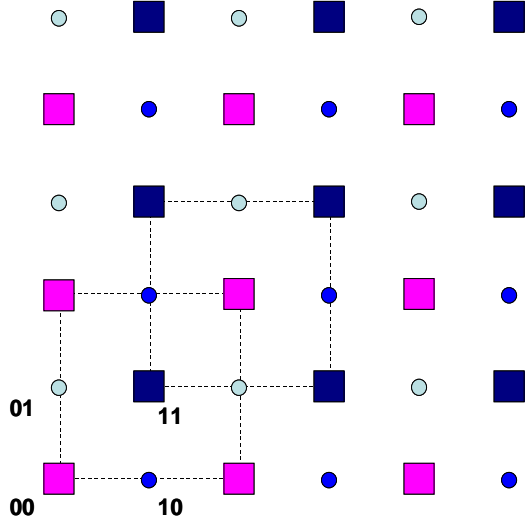,width=70mm,height=70mm}
\end{center}
\caption{Example of Construction A and set partitioning of
$\mathbb{Z}^2$ \label{fig:setpartitioning2}}
\end{figure}

As an example, let us consider a two-dimensional integer lattice
$\mathbb{Z}^2$, depicted in Fig.~\ref{fig:setpartitioning2}. In such
a lattice, the {\em checkerboard lattice} ${D}_2$ is a sublattice of
$\mathbb{Z}^2$ containing all integer vectors $(x,y)$ such that
$x+y$ is even. Using the {\em repetition code} of length two
~$C=\{(00),(11)\}$ we write
\[
D_2 = 2\mathbb{Z}^2 + C = [2\mathbb{Z}^2 + (00)] \bigcup ~
[2\mathbb{Z}^2 +(11)]
\]
This is illustrated in Fig.~\ref{fig:setpartitioning2}, where the
squares denote the ${D}_2$ lattice that is the union of the
$2\mathbb{Z}^2$ lattice (light squares) and its translate (dark
squares).

Similarly, given the {\em universe code} $C_0=(2,2,1) =
\{(00),(01),(10),(11)\}$, we can write
\[
\mathbb{Z}^2 = 2\mathbb{Z}^2 + C_0
\]
Given the linear code $C$, the {\em dual code} $C^\perp$ is defined
such that $C \oplus C^\perp = C_0$, i.e., all the binary sums of a
codeword from $C$ with a codeword from $C^\perp$ yield all the
universe codewords. In our example, $C=\{(00),(11)\}$  has a dual
code $C^\perp=\{(00),(01)\}$.

Linearity of the codes is related to the additive group structure
and enables to interpret codes and subcodes as groups and subgroups.
In turn, this lets us define a {\em quotient group} between a code
and its subcode.

For example given that $C\subset C_0$ we can write the quotient
group as the set of two cosets of the subgroup $C$, i.e.,
$C_0/C=\{C+(00),C+(01)\}$.

A well known property of abelian groups tells us that the quotient
group has itself a group structure. The quotient group operation
$\oplus$ between two cosets is defined as $(C + {\bf c}_1) \oplus (C
+ {\bf c}_2) = C + ({\bf c}_1 + {\bf c}_2)$. This implies  that the
quotient group is isomorphic to the so called {\em quotient code}
denoted by $[C_0/C]$ and defined as the set of all the {\em coset
leaders}. If $C_0$ is the universe code then the quotient code
coincides with the dual code, i.e.,
\begin{equation} \label{eq:dualcode}
[C_0/C] = C^\perp
\end{equation}
In our example $[C_0/C]=\{(00),(01)\}$.

Let us consider a lattice $\Lambda_0$ and sublattice
$\Lambda\subset\Lambda_0$. Thanks to the group structure of
lattices, we can define the {\em quotient lattice}
$\Lambda_0/\Lambda$ as the set of all distinct translates (or
cosets) of $\Lambda$, i.e.,
\begin{equation}
\Lambda_0/\Lambda = \{\Lambda+{\bf x}_i\}
\end{equation}
where ${\bf x}_i$ are the translation vectors or coset leaders. Let
$[\Lambda_0/\Lambda]$ denote the set of all the coset leaders then
we write
\begin{equation}
\Lambda_0 = \Lambda +[\Lambda_0/\Lambda]
\end{equation}

If $C_0$ and $C$ are the corresponding binary codes defined by
Construction A, we have the following group isomorphism
\begin{equation}
C_0/C \sim \Lambda_0/\Lambda
\end{equation}
Note that the quotient group defines a {\em partition} of $C_0$ into
disjoint cosets of the same size $N_c =|C_0/C|$, where $|\cdot|$
denotes the cardinality of the set. Thanks to the above isomorphism,
the {\em index} of the sublattice in the lattice is finite, i.e.,
$|\Lambda_0/\Lambda| = N_c$. Considering the fundamental volume of a
lattices defined as  ${\rm vol}(\Lambda) = \det(MM^T)^{1/2}$, where
$M$ is the lattice generator matrix, we
 have ${\rm vol}(\Lambda)/{\rm vol}(\Lambda_0)=N_c.$

Consider the sequence of nested lattices $2\mathbb{Z}^n \subseteq
\Lambda \subset \Lambda_0 \subseteq \mathbb{Z}^n$. Each coset of the
quotient lattice can be identified by a coset leader which is
related to the quotient code as follows
\begin{equation}
[\Lambda_0/\Lambda] \!\!\!\mod 2 = [C_0/C] \mbox{~~~~and~~~~}
[\Lambda/2\mathbb{Z}^n] \!\!\!\mod 2 = C
\end{equation}
This is due to the fact that the lattice $2\mathbb{Z}^n$  is
obtained by Construction A with the $(n,0)$ code, containing only
the all zero codeword. The partitions of the basic lattice
$\Lambda_0$ can be written as
\begin{equation} \label{eq:general_partition}
\Lambda_0 = \Lambda + [\Lambda_0/\Lambda]= \Lambda +[C_0/C]
\end{equation}

\begin{figure}
\begin{center}
\setlength{\unitlength}{0.8mm}
\begin{picture}(150,80)(0,-15) \thicklines
\put(75,60){\vector(-3,-2){45}}
\put(75,60){\vector(3,-2){45}} 
\put(30,30){\vector(2,-3){20}} 
\put(30,30){\vector(-2,-3){20}}
\put(120,30){\vector(-2,-3){20}} 
\put(120,30){\vector(2,-3){20}} 
\put(81,58){{$\mathbb{Z}^2$}}
\put(10,-7){00} \put(45,-7){11}\put(95,-7){01}\put(135,-7){10}
\put(123,30){{$\mathbb{Z}^2+[\mathbb{Z}^2/D_2]$}}
\put(37,-18){{$2\mathbb{Z}^2+[\mathbb{Z}^2/D_2]+
[D_2/2\mathbb{Z}^2]$}}
\end{picture}
\end{center}
\caption{The two-way partition tree of $\mathbb{Z}^2$
\label{fig:partitiontree2}}
\end{figure}
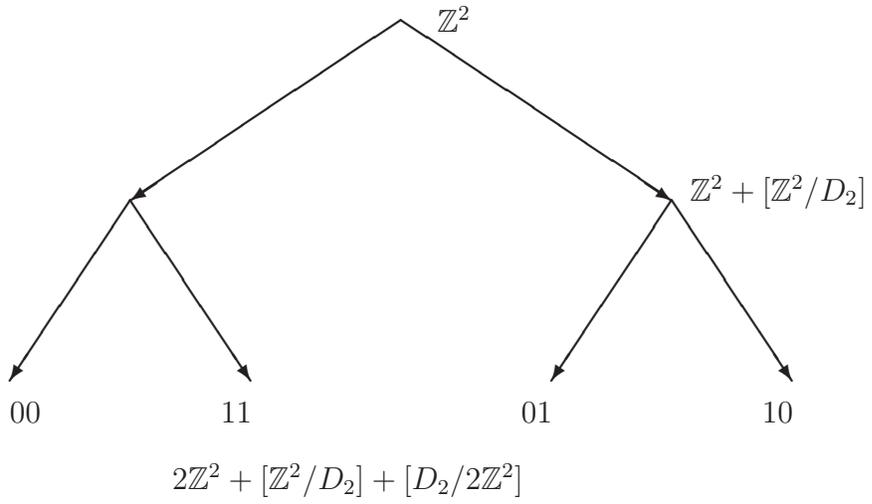

In our example, we first partition $\mathbb{Z}^2$ into two cosets:
the sublattice $D_2$  and its translate $D_2+(01)$ (squares and
circles in Fig.~\ref{fig:setpartitioning2}, respectively).
\[
\mathbb{Z}^2 = D_2 + C^\perp= D_2 + [\mathbb{Z}^2/D_2]
\]
The number of partitions equals to the index of the sublattice $D_2$
in $\mathbb{Z}^2$ and equals $N_c = |C^\perp| = 2$. We can further
partition each coset by partitioning $D_2$ into two cosets. The
sequence of nested lattices $\mathbb{Z}^2 \supset D_2 \supset
2\mathbb{Z}^2$ induces a  partition chain
\[
\mathbb{Z}^2 =  2\mathbb{Z}^2 +C^\perp +C =  2\mathbb{Z}^2 +
[\mathbb{Z}^2/D_2] + [D_2/2\mathbb{Z}^2]
\]
which can be represented by the  {\em two level binary partition
tree} in Fig.~\ref{fig:partitiontree2}.

We observe how Construction A yields a simple bit labeling of a
finite constellation ${\cal S}=\Lambda \cap {\cal B}$ carved from
the infinite lattice with shaping region ${\cal B}$. In particular,
since $\Lambda = 2\mathbb{Z}^n + C$, with $C=(n,k)$ generated by
code generator matrix $G$, the constellation points are written as
${\bf x} = 2{\bf u} + {\bf c}$, with $2{\bf u}\in 2\mathbb{Z}^n \cap
{\cal B}$ and ${\bf c}\in C$.

In order to label the constellation points ${\bf x}$, we form the
bit label vector  ${\bf b}$ as the concatenation of two parts ${\bf
b}_2$ and ${\bf b}_3$, i.e, ${\bf b}=({\bf b}_2,{\bf b}_3)$. The
first part ${\bf b}_2$ has $k$ bits and indexes the codeword ${\bf
c} = {\bf b}_2 G$. The second part ${\bf b}_3$ labels the integer
vectors ${\bf u}$, such that $2{\bf u}+{\bf c} \in {\cal B}$. Note
that the number of bits in ${\bf b}_3$ depends on the size of ${\cal
B}$.
When ${\cal B}$ has a cubic shape, we can apply a Gray labeling to
each component of ${\bf u}$.

\begin{figure}
\begin{center}
  \psfig{file=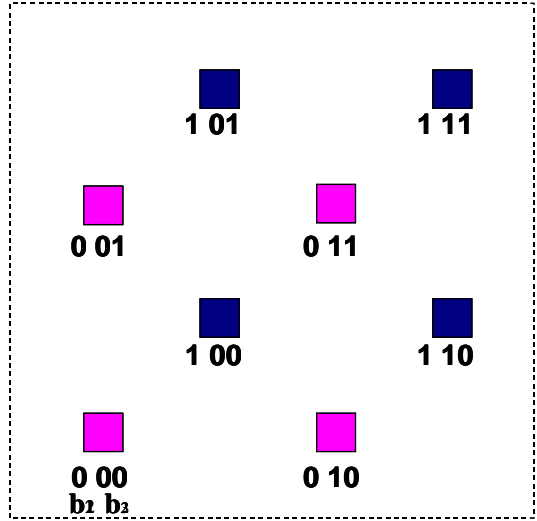,width=70mm,height=70mm}
\end{center}
\caption{Labeling the finite constellation carved from $D_2$
\label{fig:labelingD2} }
\end{figure}

For example, Fig.~\ref{fig:labelingD2} shows the labeling of an 8
point constellation carved from $D_2$, where one bit is used to
select one on the two codewords (00) and (11), while the other two
bits to select one of the four points in $2\mathbb{Z}^2 \cap {\cal
B}$.

Finally, we consider the labeling of the entire finite constellation
carved from $\Lambda_0\subseteq \mathbb{Z}^n$. In order to follow
the partition into cosets induced by $\Lambda\subset\Lambda_0$, we
use (\ref{eq:general_partition}) to get
\begin{equation}
\Lambda_0 = \Lambda + [\Lambda_0/\Lambda] + [\Lambda/2\mathbb{Z}^n]=
2\mathbb{Z}^n + [C_0/C] + C
\end{equation}
In particular, we add ${\bf b}_1$ information bits, which are used
to label the codewords of the quotient code $[C_0/C]$. So the final
bit label is ${\bf b}=({\bf b}_1, {\bf b}_2, {\bf b}_3)$.

\begin{figure}
\begin{center}
  \psfig{file=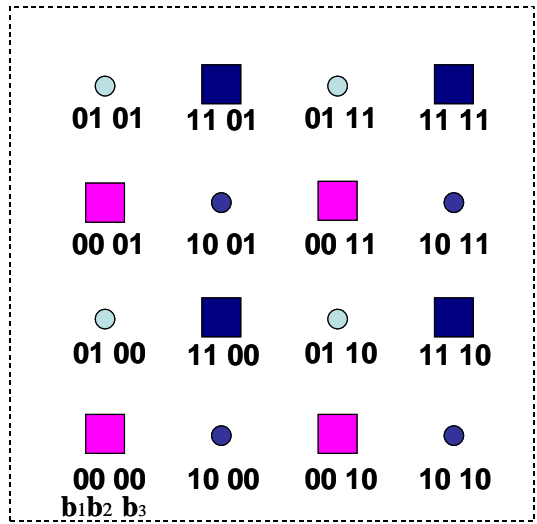,width=70mm,height=70mm}
\end{center}
\caption{Labeling the finite constellation carved from
$\mathbb{Z}^2$ using the two level set partitioning
\label{fig:labeling16QAM}}
\end{figure}

Fig.~\ref{fig:labeling16QAM} shows the labeling of the 16--QAM
obtained by set partitioning corresponding to
Fig.~\ref{fig:setpartitioning2}. The extra bit ${\bf b}_1$ selects
one of the two codewords of the dual code (00) and (01), while ${\bf
b}_2$ and ${\bf b}_3$ are the same as in Fig.~\ref{fig:labelingD2}.
This labeling technique was first proposed by Ungerboeck and we can
observe how the overall labeling is {\em not} a Gray labeling of the
16--QAM.

\bibliographystyle{IEEE}

{\small
\begin{thebibliography}{10}
\itemsep=1mm
\parskip=0mm

\bibitem{Tarokh98}
V.~Tarokh, N.~Seshadri and A.~R.~Calderbank, ``Space-Time Codes for
High Data Rate Wireless Communications: Performance Criterion and
Code Construction,'' {\em IEEE Transactions on Information Theory},
vol.~44, no.~2, pp.~744--765, 1998.

\bibitem{Alamouti}
S.~M.~Alamouti,
\newblock ``A simple transmit diversity technique for
wireless communications,''
\newblock {\em IEEE Journals of Selected Areas on Communications},
vol.~16, no.~8, pp.~1451--1458, Oct.~1998.

\bibitem{Tse03}
Lizhong Zheng and  D.N.C.~Tse, ``Diversity and multiplexing: a
fundamental tradeoff in multiple-antenna channels,'' {\em IEEE
Transactions on Information Theory}, vol.~49, no.~5, p.~1073--1096,
May 2003.

\bibitem{STBCTarokh}
V.~Tarokh, H.~Jafarkhani and A.~R.~Calderbank, ``Space-time block
codes from orthogonal designs,'' {\em IEEE Transactions on
Information Theory}, vol.~45, no.~5, pp.~1456--1467, July 1999.


\bibitem{DAST1}
M.~O.~Damen, K.~Abed-Meraim, and J.-C.~Belfiore, ``Diagonal
algebraic space-time block codes,'' {\em IEEE Transactions on
Information Theory}, vol. 48, pp. 628-636, Mar. 2002.

\bibitem{DAST2}
M.~O.~Damen, K.~Abed-Meraim, and J.-C.~Belfiore,
 ``Transmit diversity using rotated constellations with Hadamard transform,''
 {\em IEEE Proc. 2000 Symp. Adaptive Systems for Signal Processing, Communications, and Control}, AB, Canada, pp. 396-401, Oct. 2000.

\bibitem{UniversalSTC03}
H.~El~Gamal and M.~O.~Damen,
 ``Universal space-time codes,''
 {\em IEEE Transactions on Information Theory}, vol. 49, no. 5, pp. 1097-1119, May 2003.

\bibitem{Sethuraman03}
B.~A.~Sethuraman, B.~S.~Rajan, and V.~Shashidhar,
 ``Full-diversity, high-rate space-time block codes from division algebras,''
 {\em IEEE  Transactions on Information Theory}, vol. 49, pp. 2596-2616, Oct. 2003.

\bibitem{Golden05}
J.-C.~Belfiore, G.~Rekaya, and E.~Viterbo, ``The Golden Code: A $2
\times 2$ full-rate space-time code with non-vanishing
determinants,'' {\em IEEE Transactions on Information Theory},
vol.~51, no.~4, pp.~1432--1436, Apr.~2005.

\bibitem{Elia05}
P.~Elia,  K.R.~Kumar, S.A.~Pawar, P.V.~Kumar, Hsiao-feng Lu,
``Explicit space-time codes that achieve the diversity-multiplexing
gain tradeoff,'' {\em International Symposium on Information Theory,
ISIT 2005}, p.~896--900, Adelaide, Australia, 4-9th Sept. 2005.

\bibitem{Viterbo99}
E.~Viterbo and J.~Boutros, ``A Universal Lattice Code Decoder for
Fading Channels,'' {\em IEEE Transactions on Information Theory},
vol.~45, no.~5, pp.~1639--1642, July 1999.

\bibitem{Jafarkhani03}
H.~Jafarkhani and N.~Seshadri, ``Super-orthogonal space-time trellis
codes,''  {\em IEEE Transactions on Information Theory}, vol.~49,
no.~4, pp.~937--950, April 2003.

\bibitem{Calderbank}
A.~R.~Calderbank and N.~J.~Sloane, ``New trellis codes based on
lattices and cosets,'' {\em IEEE Transactions on Information
Theory}, vol.~33, no.~2, pp.~177--195, Mar.~1987.

\bibitem{Ungerboeck}
G.~Ungerboeck, ``Trellis Coded Modulation with Redundant Signal Sets.
Part II: State of the Art,'' {\em IEEE Communications Magazine},
vol.~25, n.~2, pp.~12--21, Feb.~1987.

\bibitem{Forney1}
G.~D.~Forney~Jr., ``Coset codes. I. Introduction and geometrical
classification'' {\em IEEE Transactions on Information Theory,} vol.
34, Sept. 1988, pp. 1123--1151.

\bibitem{Forney2}
G.~D.~Forney~Jr., ``Coset codes. II. Binary lattices and related
codes,'' {\em IEEE Transactions on Information Theory,} vol.~34,
Sept.~1988, pp.~1152--1187.

\bibitem{Foundation04}
F.~Oggier and E.~Viterbo, ``Algebraic number theory and code
design for Rayleigh fading channels,'' {\em Foundations and Trends
in Communications and Information Theory}, vol.~1, pp.~333-415,
2004.

\bibitem{David05}
D.~Champion, J.-C.~Belfiore, G.~Rekaya and E.~Viterbo,
``Partitionning the Golden Code: A framework to the design of
Space-Time coded modulation,'' {\em Canadian Workshop on
Information Theory}, 2005.

\bibitem{Biglieri05}
E.~Biglieri, {\em Coding for wireless channels}, Springer, New York,
2005.

\bibitem{groupsbook}
J.~Rotman, {\em An introduction to the theory of groups}, Springer,
New York, 1994.

\bibitem{Spherepacking}
J.~H.~Conway and N.~J.~A.~Sloane,
``Sphere Packings, Lattices and Groups,''
{Springer-Verlag, New York}, 1992.

\bibitem{matrixbook} H. L\"utkepolhl , {\em Handbook of Matrices},
Chichester, England, John Wiley \& Sons Ltd., 1996.

\bibitem{MacWilly}
F.J. MacWilliams and N.J.A. Sloane, {\em The Theory of Error-Correcting Codes,}
North-Holland, Amsterdam, 1977.
\end{thebibliography}
}

\begin{figure}
\begin{center}
\setlength{\unitlength}{0.8mm}
\begin{picture}(150,60) \thicklines%
\put(75,60){\vector(-3,-2){45}}\put(75,60){\vector(-1,-2){15}}
\put(75,60){\vector(3,-2){45}} \put(75,60){\vector(1,-2){15}}
\put(30,30){\vector(0,-1){30}} \put(30,30){\vector(-1,-3){10}}
\put(30,30){\vector(-2,-3){20}}\put(30,30){\vector(-1,-1){30}}
\put(60,30){\vector(0,-1){30}} \put(60,30){\vector(-1,-3){10}}
\put(60,30){\vector(-2,-3){20}}\put(60,30){\vector(1,-3){10}}
\put(90,30){\vector(0,-1){30}} \put(90,30){\vector(-1,-3){10}}
\put(90,30){\vector(2,-3){20}} \put(90,30){\vector(1,-3){10}}
\put(120,30){\vector(0,-1){30}} \put(120,30){\vector(1,-1){30}}
\put(120,30){\vector(2,-3){20}} \put(120,30){\vector(1,-3){10}}
\put(45,45){$\alpha_1$} \put(8,15){$\alpha_2$}
\put(81,58){{$\Lambda_{\ell_0}$}} \put(45,35){0}\put(66,35){1}
\put(82,35){2} \put(102,35){3}
\put(30,-7){{$\Lambda_{\ell_0+2}+[\Lambda_{\ell_0}/\Lambda_{\ell_0+1}]+
[\Lambda_{\ell_0+1}/\Lambda_{\ell_0+2}]$}}
\end{picture}
\end{center}
\caption{Two level ($\ell=2$) partition tree of $\Lambda_{\ell_0}$
into 16 cosets of $\Lambda_{\ell_0+2}$.} \label{Fig:setpartition}
\end{figure}
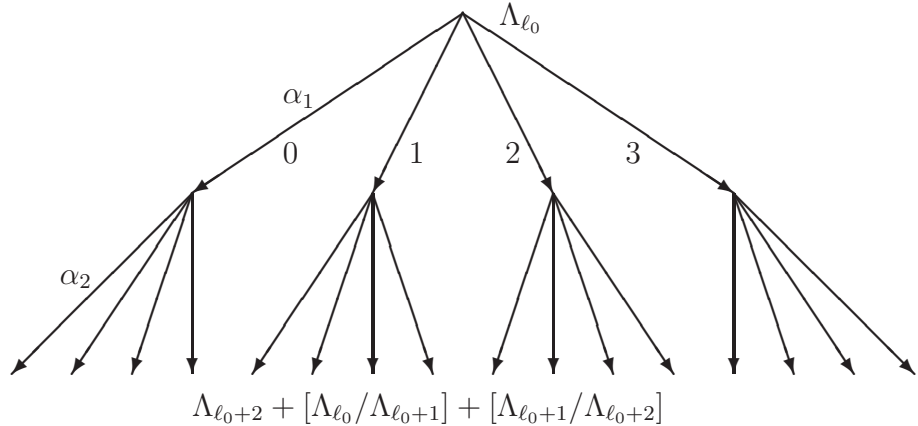

\newpage
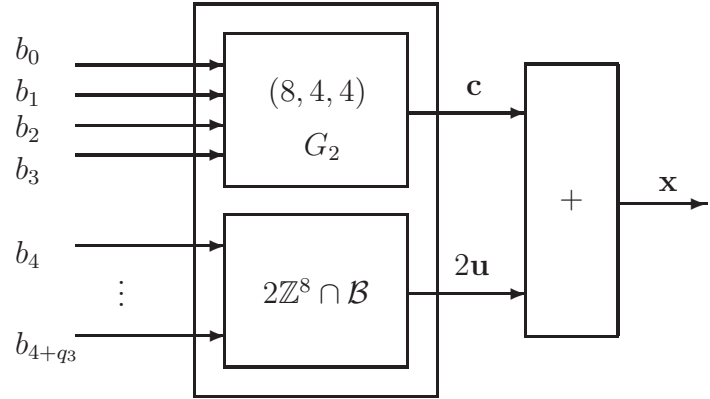
\begin{figure}[t]
\begin{center}
\setlength{\unitlength}{0.8mm}
\begin{picture}(125,65)
\thicklines%
\put(15,10){\vector(1,0){25}} \put(15,25){\vector(1,0){25}}
\put(15,40){\vector(1,0){25}} \put(15,45){\vector(1,0){25}}
\put(15,50){\vector(1,0){25}} \put(15,55){\vector(1,0){25}}
\put(70,17){\vector(1,0){20}} \put(70,47){\vector(1,0){20}}
\put(105,32){\vector(1,0){15}}
\put(40,5){\framebox(30,25){$2\mathbb{Z}^8\cap{\cal B}$}}
\put(40,35){\framebox(30,25)} \put(53,40){$G_2$}
\put(47,49){$(8,4,4)$} \put(35,0){\framebox(40,65)}
\put(90,10){\framebox(15,45){${\bf +}$}}
\put(80,50){$\mathbf{c}$}\put(78,20){$2\mathbf{u}$}
\put(112,34){$\mathbf{x}$}\put(22,15){$\vdots$} \put(5,56){$b_0$}
\put(5,49){$b_1$} \put(5,43){$b_2$} \put(5,36){$b_3$}
\put(5,22){$b_4$} \put(5,7){$b_{4+q_3}$}
\end{picture}
\end{center}
\caption{The $E_8$ encoder structure resulting in a ${\cal B}$
shaped finite constellation.} \label{Fig:E8encoder}
\end{figure}
\newpage
\begin{figure}[t]
\begin{center}
\psfig{file=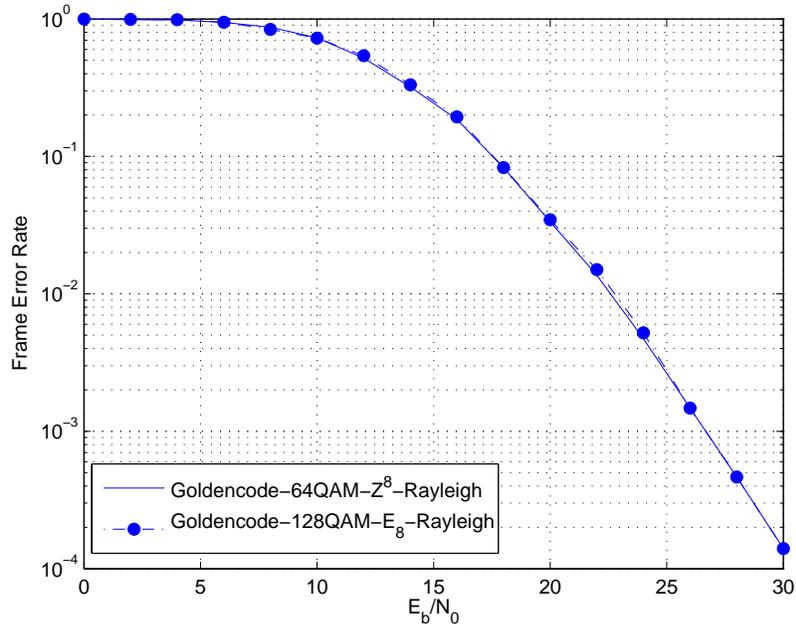,width=120mm,height=90mm}
\end{center}
\caption{Performance of $\mathbb{Z}^{8}$ Golden code with 64-QAM and
$E_8$ Golden subcode with 128-QAM (12bpcu).}
\label{Fig:PerfE8Gold128}
\end{figure}
\newpage
\begin{figure}[t]
\begin{center}
\psfig{file=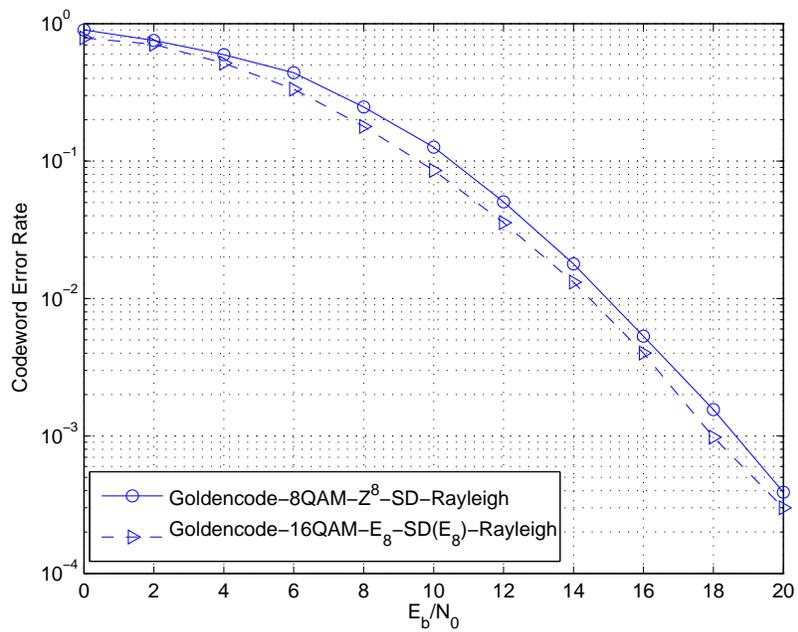,width=120mm,height=90mm}
\end{center}
\caption{Performance of $\mathbb{Z}^{8}$ Golden code with 8-QAM and
$E_8$ Golden subcode with 16-QAM (6bpcu).} \label{Fig:PerfE8Gold16}
\end{figure}
\newpage
\begin{figure}[t]
\begin{center}
\psfig{file=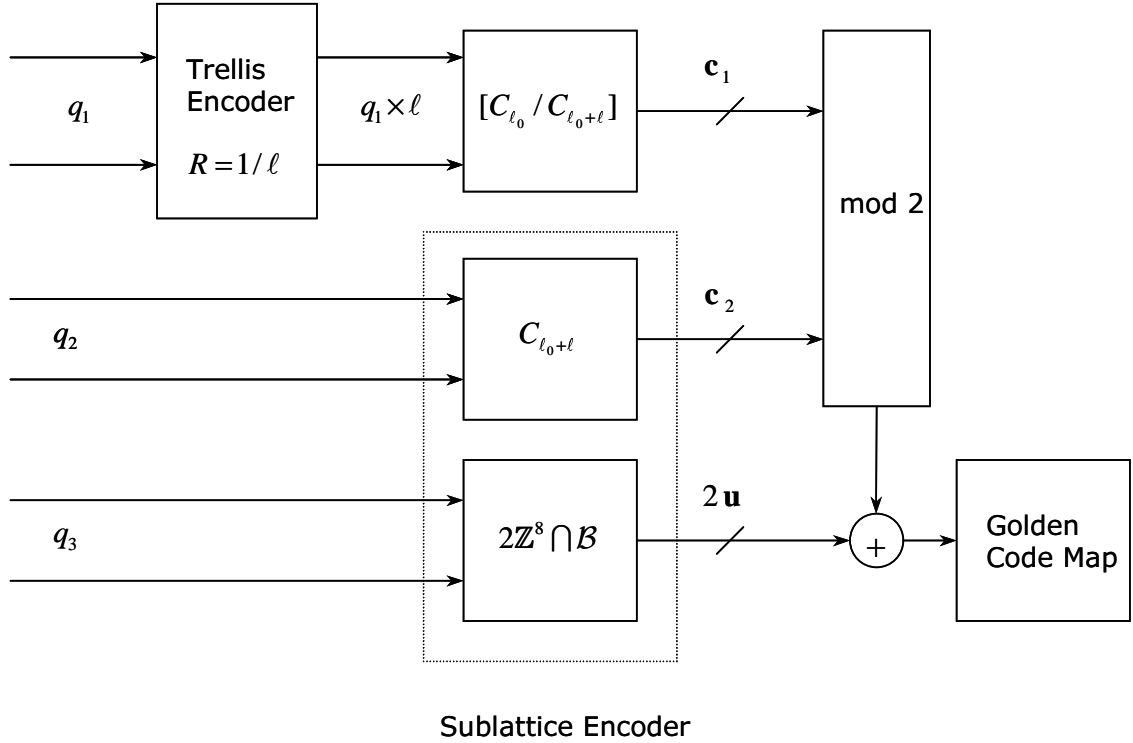,width=150mm,height=100mm}
\end{center}
\caption{General encoder structure of the concatenated scheme.}
\label{Fig:systemmodel}
\end{figure}
\newpage
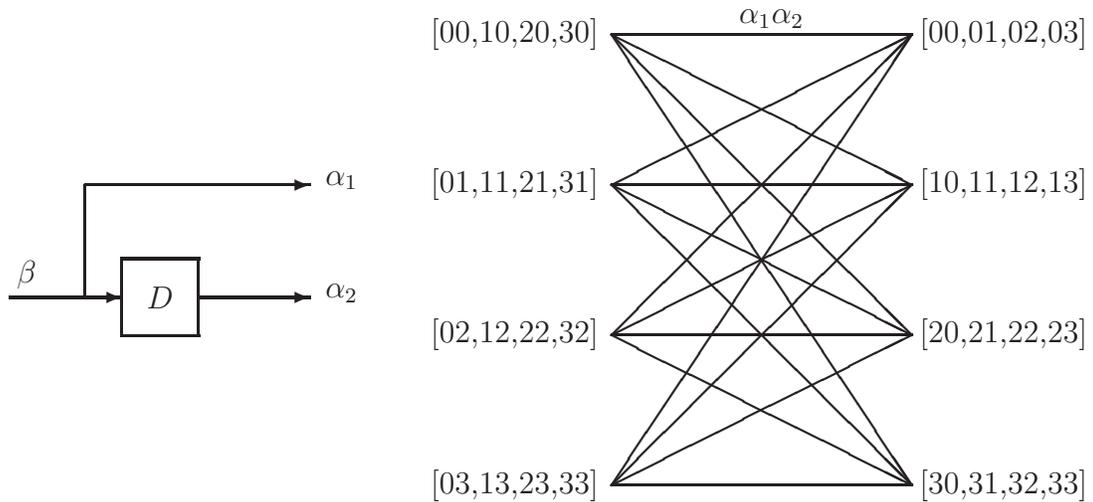
\begin{figure}[t]
\begin{center}
\setlength{\unitlength}{1mm}
\begin{picture}(140,70)
\thicklines%
\put(-5,20){\begin{picture}(60,30) %
\put(5,10){\vector(1,0){15}} \put(20,5){\framebox(10,10){$D$}}
\put(30,10){\vector(1,0){15}} \put(15,25){\vector(1,0){30}}
\put(15,10){\line(0,1){15}} \put(6,12){$\beta$}
\put(47,25){$\alpha_1$} \put(47,10){$\alpha_2$}
\end{picture}}%
\put(75,0){\begin{picture}(65,70) %
\put(5,65){\line(1,0){40}}  \put(5,65){\line(2,-1){40}}
\put(5,65){\line(1,-1){40}} \put(5,65){\line(2,-3){40}}
\put(5,45){\line(1,0){40}}  \put(5,45){\line(2,-1){40}}
\put(5,45){\line(1,-1){40}} \put(5,45){\line(2,1){40}}
\put(5,25){\line(1,0){40}}  \put(5,25){\line(2,-1){40}}
\put(5,25){\line(1,1){40}}  \put(5,25){\line(2,1){40}}
\put(5,5){\line(1,0){40}}   \put(5,5){\line(2,3){40}}
\put(5,5){\line(1,1){40}}   \put(5,5){\line(2,1){40}}
\put(-19,64){[00,10,20,30]} \put(-19,44){[01,11,21,31]}
\put(-19,24){[02,12,22,32]} \put(-19,4){[03,13,23,33]}
\put(46,64){[00,01,02,03]} \put(46,44){[10,11,12,13]}
\put(46,24){[20,21,22,23]} \put(46,4){[30,31,32,33]}
\put(22,66.5){$\alpha_1\alpha_2$}
\end{picture}}%
\end{picture}
\end{center}
\caption{The 4-state encoder with $g_1(D)=1$ and $g_2(D)=D$ and corresponding
trellis diagram. Labels on the left are outgoing from each state clockwise, labels
on the right are incoming counterclockwise.}
\label{Fig:trellis4states}
\end{figure}

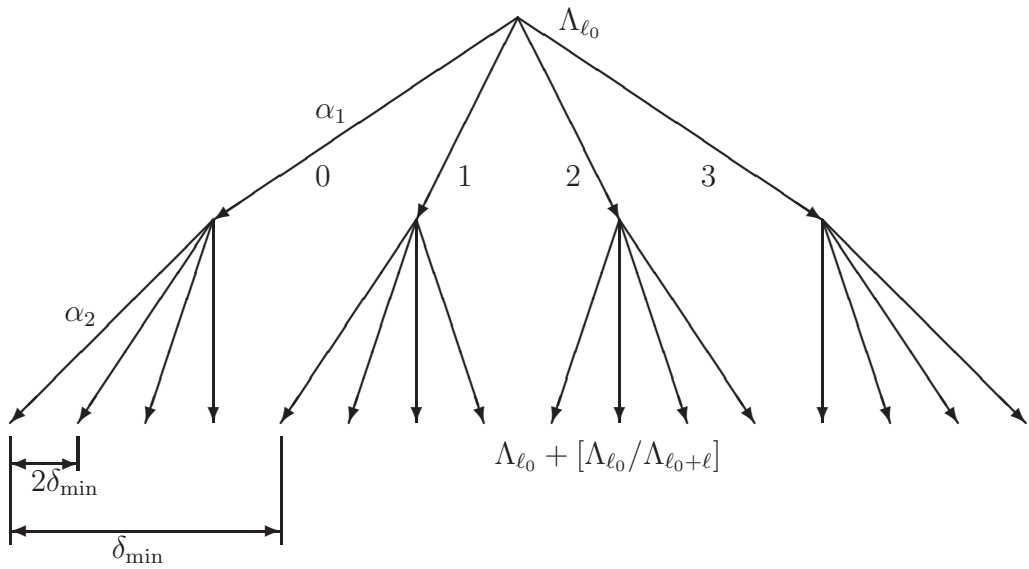
\begin{figure}[t]
\begin{center}
\setlength{\unitlength}{0.9mm}
\begin{picture}(150,70)(0,-20) \thicklines%
\put(75,60){\vector(-3,-2){45}}\put(75,60){\vector(-1,-2){15}}
\put(75,60){\vector(3,-2){45}} \put(75,60){\vector(1,-2){15}}
\put(30,30){\vector(0,-1){30}} \put(30,30){\vector(-1,-3){10}}
\put(30,30){\vector(-2,-3){20}}\put(30,30){\vector(-1,-1){30}}
\put(60,30){\vector(0,-1){30}} \put(60,30){\vector(-1,-3){10}}
\put(60,30){\vector(-2,-3){20}}\put(60,30){\vector(1,-3){10}}
\put(90,30){\vector(0,-1){30}} \put(90,30){\vector(-1,-3){10}}
\put(90,30){\vector(2,-3){20}} \put(90,30){\vector(1,-3){10}}
\put(120,30){\vector(0,-1){30}} \put(120,30){\vector(1,-1){30}}
\put(120,30){\vector(2,-3){20}} \put(120,30){\vector(1,-3){10}}
\put(45,45){{$\alpha_1$}} \put(8,15){{$\alpha_2$}}
\put(81,58){{$\Lambda_{\ell_0}$}} \put(45,35){0}\put(66,35){1}
\put(82,35){2} \put(102,35){3}
\put(0,-6){\vector(1,0){10}}\put(10,-6){\vector(-1,0){10}}
\put(3,-10){{$2\delta_{\min}$}}
\put(0,-16){\vector(1,0){40}}\put(40,-16){\vector(-1,0){40}}
\put(15,-20){{$\delta_{\min}$}}
\put(0,-2){\line(0,-1){16}}\put(10,-2){\line(0,-1){5}}\put(40,-2){\line(0,-1){16}}
\put(70,-6){{
$\Lambda_{\ell_0}+[\Lambda_{\ell_0}/\Lambda_{\ell_0+\ell}]$}}
\end{picture}
\end{center}
\caption{Inter coset distances for a two level partition tree}
\label{Fig:intercoset}
\end{figure}

\begin{figure}[t]
\begin{center}
\setlength{\unitlength}{1mm} \thicklines
\begin{picture}(90,50)(-5,5) %
\put(0,25){\vector(1,0){20}} \put(20,20){\framebox(10,10){$D$}}
\put(30,25){\vector(1,0){20}} \put(50,20){\framebox(10,10){$D$}}
\put(65,10){\circle{7}}\put(64,9){$+$} \put(10,10){\vector(1,0){51}}
\put(65,25){\vector(0,-1){11}}
\put(40,35){\vector(1,0){40}}\put(69,10){\vector(1,0){10}}
\put(10,10){\line(0,1){15}} \put(10,10){\line(1,0){5}}
\put(40,25){\line(0,1){10}} \put(60,25){\line(1,0){5}}
\put(6,27){$\beta$} \put(82,32){$\alpha_1$} \put(82,10){$\alpha_2$}
\end{picture}
\end{center}
\begin{center}
\psfig{file=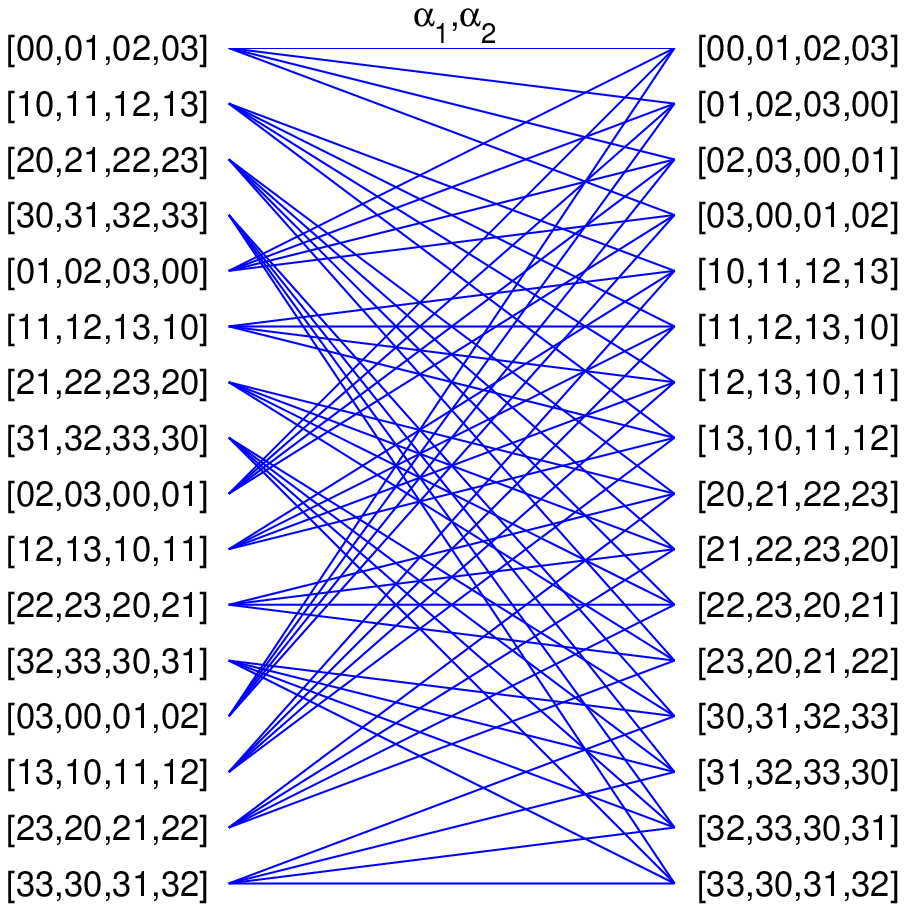,height=100mm}
\end{center}\vspace{-1.5cm}
\caption{The optimal 16 states trellis corresponding to the
generators $g_1(D)=D$ and $g_2(D)=1+D^2$. Labels on the left are
outgoing from each state clockwise, labels on the right are incoming
counterclockwise.} \label{Fig:trellis16states}
\end{figure}
\newpage
\begin{figure}[t]
\begin{center}
\psfig{file=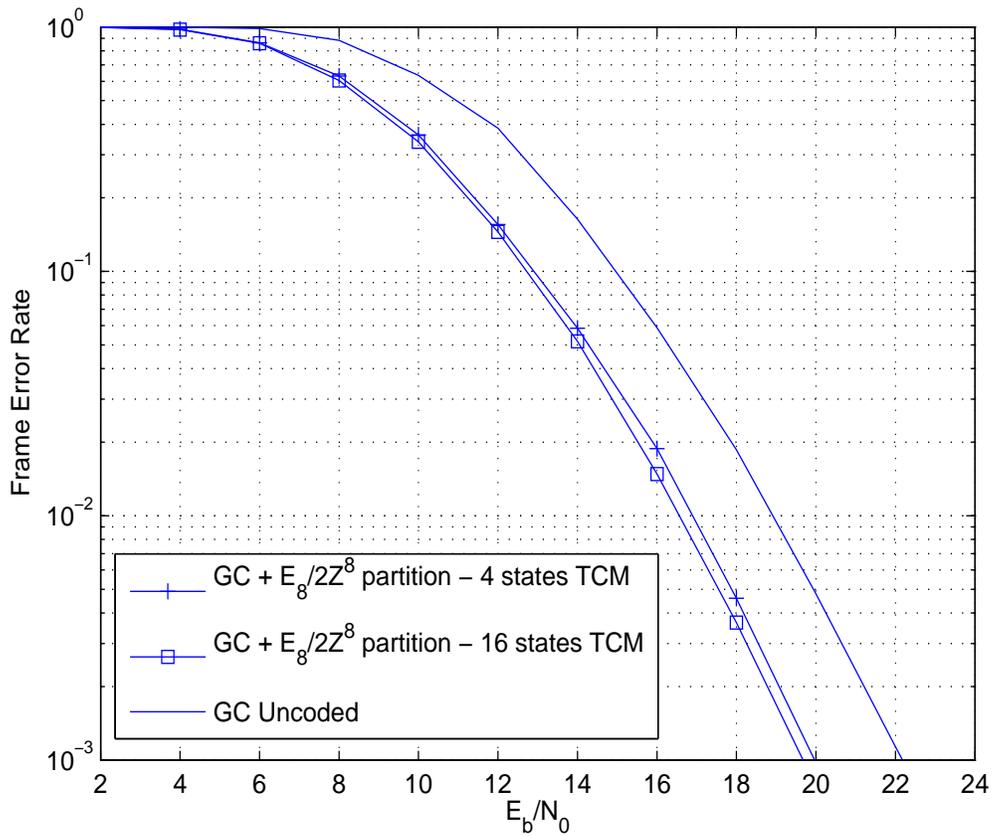,width=150mm,height=120mm}
\end{center}
\caption{Performance comparison of a 4-state trellis code using
16-QAM constellation and an uncoded transmission at the rate 5 bpcu,
$\Lambda =E_8$,  $\Lambda_{\ell} = 2\mathbb{Z}^8$, $\ell = 2$ (see
Example 1).} \label{Fig:8}
\end{figure}
\newpage
\begin{figure}[t]
\begin{center}
\psfig{file=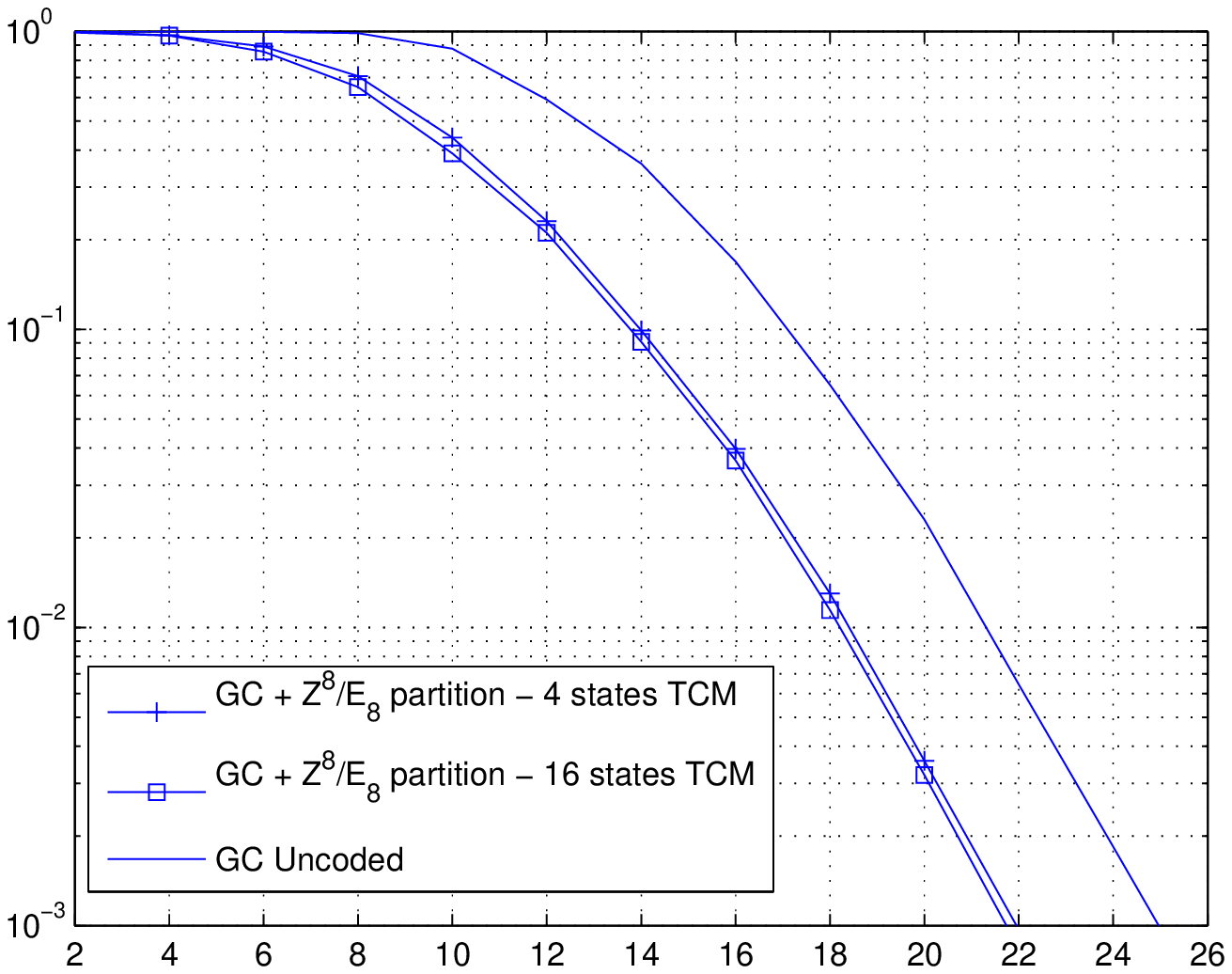,width=150mm,height=120mm}
\end{center}
\caption{Performance comparison of 4 and 16 state trellis codes
using 16-QAM constellation and an uncoded transmission at the rate
of 7 bpcu and $\Lambda = \mathbb{Z}^8, \Lambda_\ell = E_8$, $\ell
=2$ (see Example 2).} \label{Fig:9}
\end{figure}
\newpage
\begin{figure}[t]
\begin{center}
\psfig{file=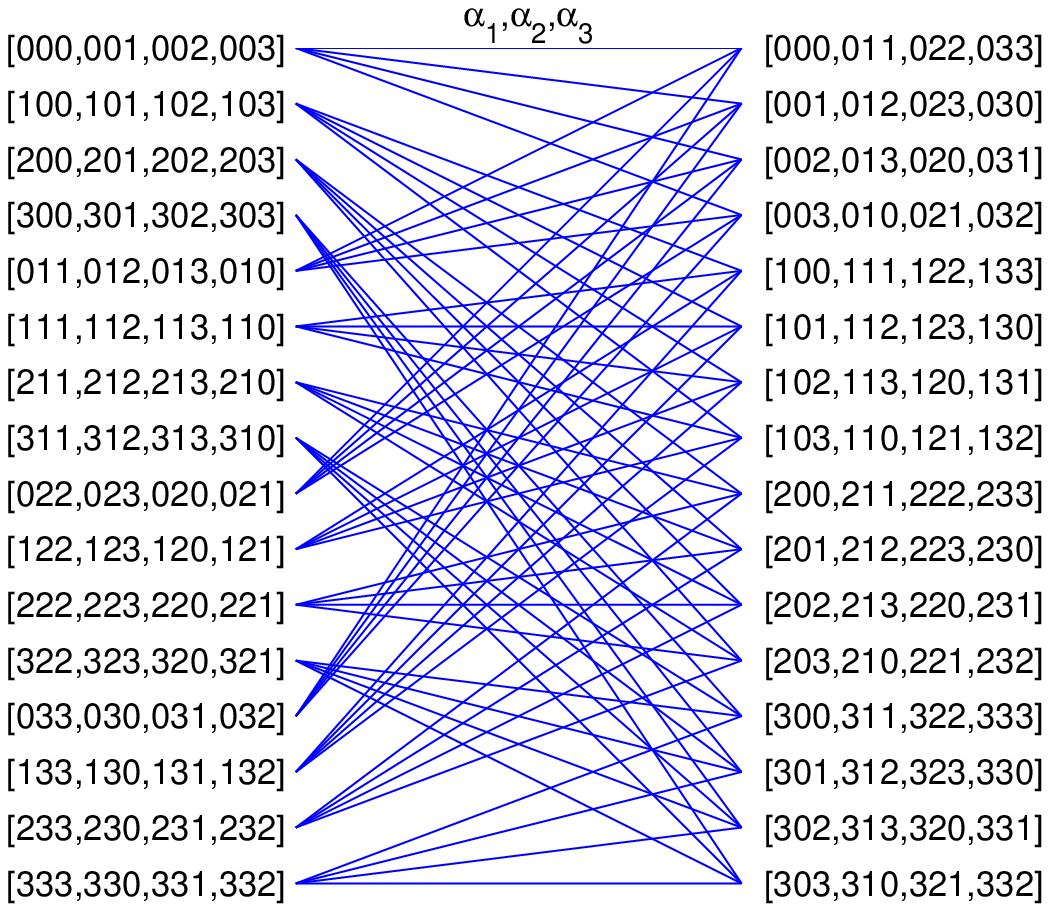,height=100mm}
\end{center}\vspace{-1.5cm}
\caption{The 16 states trellis corresponding to the generators
$g_1(D)=D$, $g_2(D)=D^2$, and $g_3(D)=1+D^2$. Labels on the left are
outgoing from each state clockwise, labels on the right are incoming
counterclockwise.} \label{Fig:trellis16statesr3}
\end{figure}
\newpage
\begin{figure}[t]
\begin{center}
\setlength{\unitlength}{1mm}
\begin{picture}(120,180)
\put(-50,-15){\psfig{file=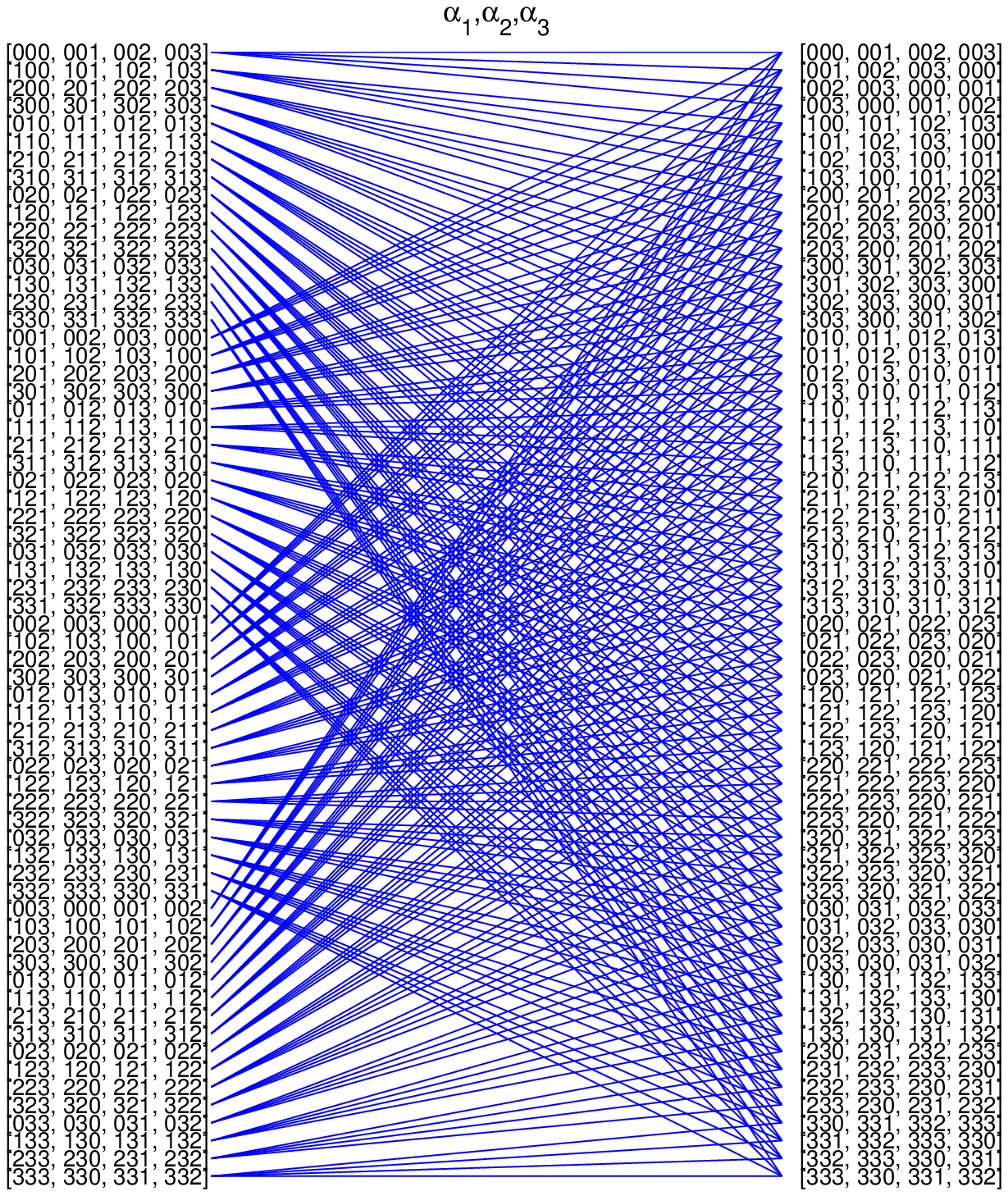,height=180mm}}
\end{picture}
\end{center}
\caption{The optimal 64 states trellis corresponding to the
generators $g_1(D)=D$, $g_2(D)=D^2$, and $g_3(D)=1+D^3$. Labels on
the left are outgoing from each state clockwise, labels on the right
are incoming counterclockwise.} \label{Fig:trellis64statesr3}
\end{figure}
\newpage
\begin{figure}[t]
\begin{center}
\psfig{file=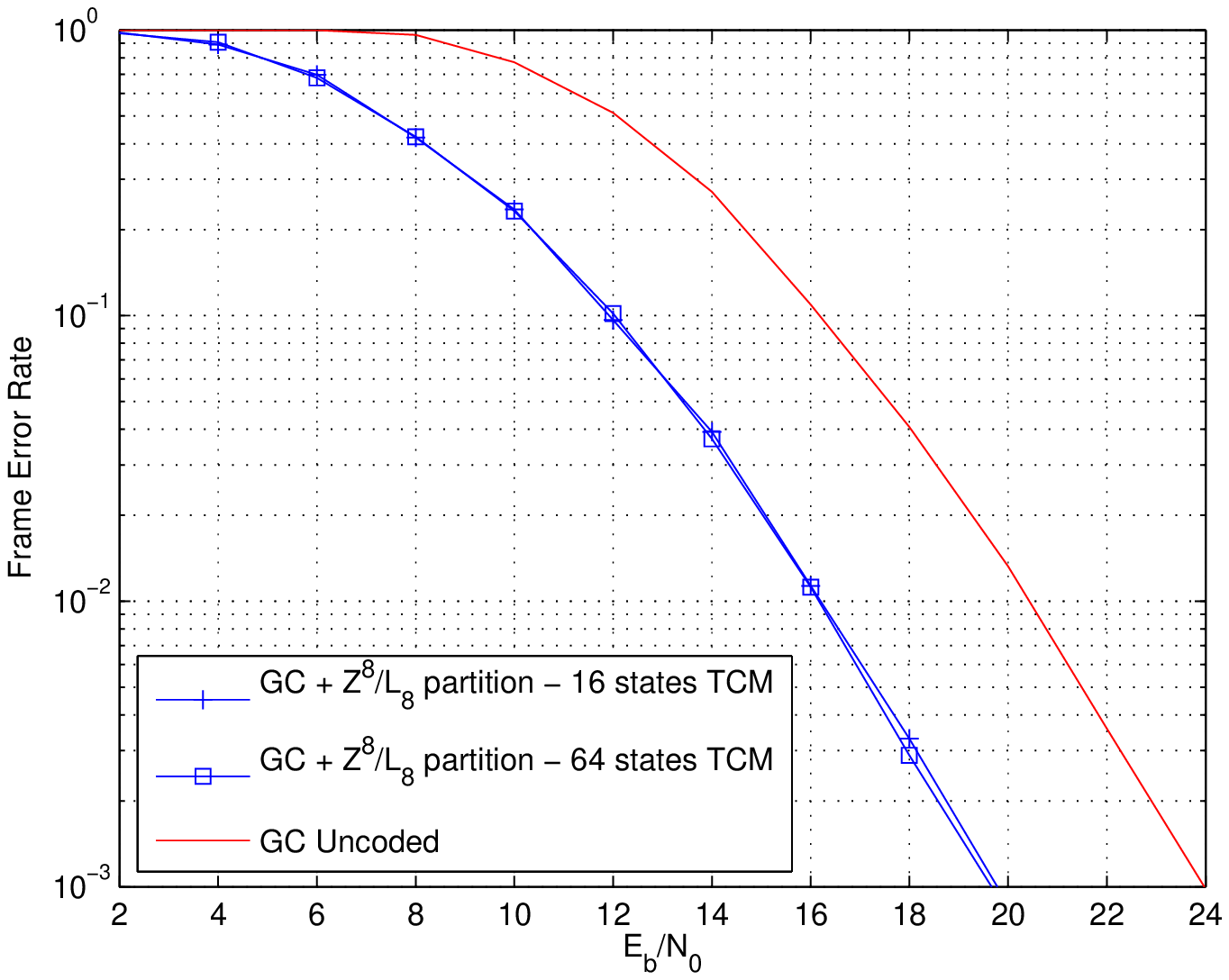,width=150mm,height=120mm}
\end{center}
\caption{Performance comparison of 16 and 64 state trellis codes
using 16-QAM constellation and an uncoded transmission at the rate
of 6 bpcu and $\Lambda = \mathbb{Z}^8, \Lambda_\ell =L_8, \ell =3
$ (see Example 3).} \label{Fig:12}
\end{figure}
\newpage
\begin{figure}[t]
\begin{center}
\psfig{file=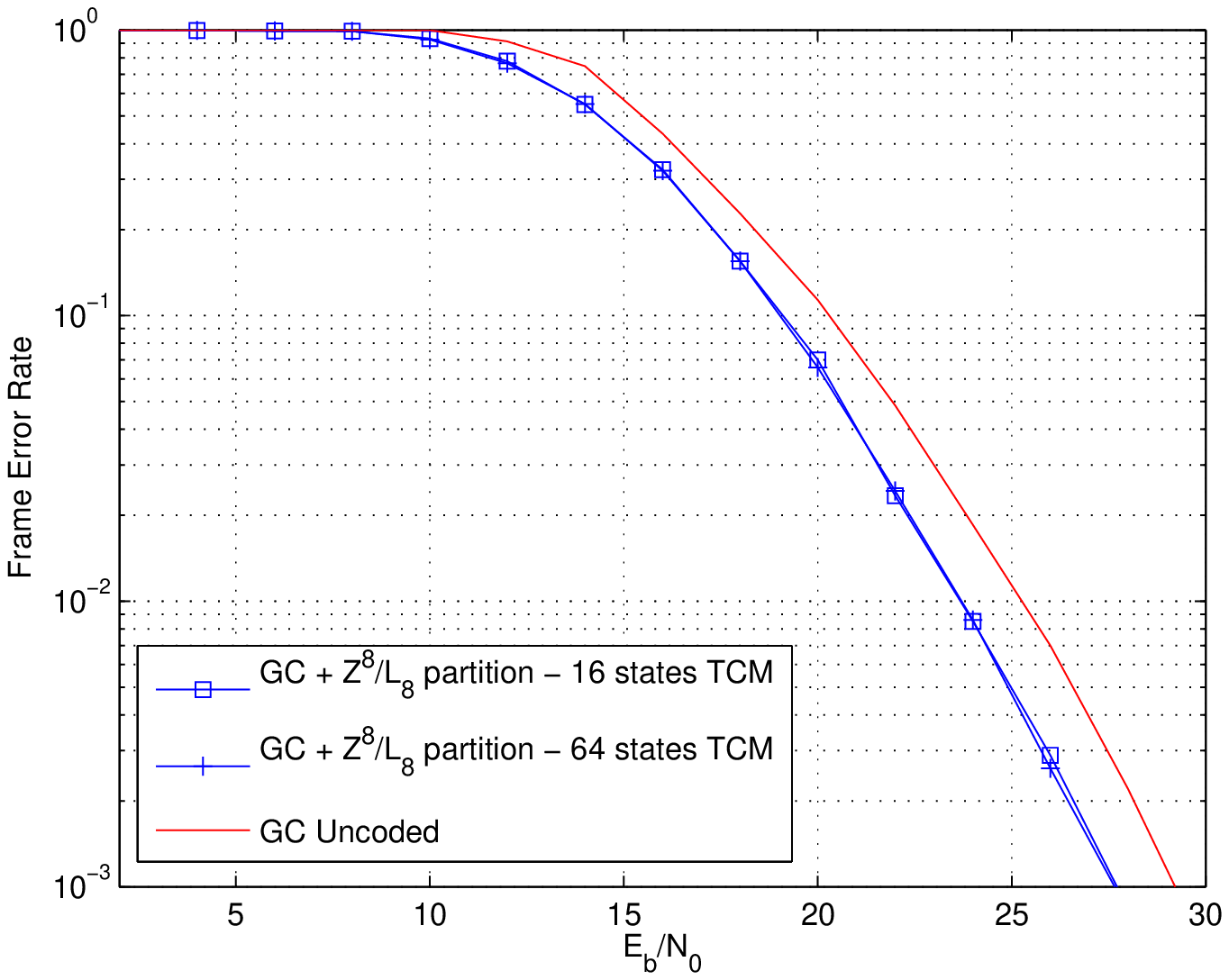,width=150mm,height=120mm}
\end{center}
\caption{Performance comparison of 16 and 64 state trellis codes
using 64-QAM constellation and an uncoded transmission at the rate
of 10 bpcu and $\Lambda = \mathbb{Z}^8, \Lambda_\ell =L_8, \ell =3
$ (see Example 4).} \label{Fig:13}
\end{figure}

\end{document}